# Edge Games: Cooperative Partner Selection in Network Cooperation Evolution

Hongqian Wu[1], Hongzhong Deng[1], Jichao Li[1], Chengxing Wu[1], Zhuoting Yu[1], Haidong Zhang[1], Gaoxin Qi[2]

[1]*College of Systems Engineering, National University of Defense Technology, Changsha 410073, China*

[2]*The Third Institute, Information Engineering University, Zhengzhou 450001, China*

**Abstract**

The phenomenon of group cooperation constitutes a fundamental mechanism underlying various social and biological systems. Complex networks provide a structural framework for group interactions, where individuals can not only obtain information from their neighbors but also choose neighbors as cooperative partners. However, traditional evolutionary game theory models, where nodes are the game players, are not convenient for directly choosing cooperative partners. Here, we exchange the roles of nodes and edges and innovatively propose the "edge game" model, using edges in complex networks as virtual game players for group games. Theoretical analysis and simulation experiments show that by configuring a synergy factor ($r$) that satisfies the "moderate cooperation" condition, a stable cooperative structure can be achieved for any network at the evolutionary equilibrium. Specifically, when there is no constraint on the number of cooperators per node, the condition for the evolution of cooperation in the network is $r > k_{max}$, where $k_{max}$ is the maximum degree of the nodes. When there is a threshold constraint, in nearest-neighbor coupled networks (with degree $k$), the condition for "moderate cooperation" is $k < r < 2k$. In heterogeneous networks, a variable synergy factor scheme is adopted, where the synergy factor for each game group ($r_x$) is defined to be proportional to the degree of the central node ($k_x$) in the group ($r_x = n\text{-}fold \cdot k_x$), "moderate cooperation" can be achieved when $1 < n\text{-}fold < 2$. If the value of $r$ exceeds the range, it may lead to "excessive cooperation" with node overload. Comparing algorithm performance and time complexity, edge games demonstrate advantages over other optimization algorithms. Simple and universal, the edge game provides a new approach to addressing multi-agent cooperation problems in the era of machine intelligence.

## 1. Introduction

The problem of collective cooperation is one of the core issues in various fields such as sociology, economics, management, biology, international relations, and artificial intelligence. It involves scenarios such as human resource sharing [1], market contract design [2], corporate team division of labor [3], cooperative hunting in animals [4], international collaboration on global issues like carbon neutrality [5,6], and multi-agent coordination in distributed self-organizing networks [7]. In these cases, rational individuals often choose group cooperation to maximize their collective benefits. However, due to the self-interested nature of individuals, the phenomenon of "free-riding" often arises in group cooperation, where individuals rely on others to bear the costs while they enjoy the benefits without contributing, destabilizing the cooperation. This creates a conflict between individual rationality and collective rationality, leading to the tragedy of the commons [8,9]. Thus, achieving a win-win situation for both individual and group interests is a crucial challenge for group cooperation.

Evolutionary game theory provides a mature theoretical framework and mathematical tools for studying group cooperation. Through repeated interactions, individuals engage in games based on their "fitness" and make strategy choices according to predefined evolutionary rules. Over time, as interactions accumulate, the strategy distribution within the group will converge towards an evolutionarily stable strategy (ESS) [10]. By adjusting the rules or parameters to study different equilibrium states, evolutionary game theory can reveal the environmental and mechanistic conditions under which stable cooperation within a group can be achieved.

Previous research has identified five typical mechanisms that can promote the emergence of cooperation: kin selection, direct reciprocity, indirect reciprocity, network reciprocity, and group selection [11]. Among these, network reciprocity can more precisely model the non-uniform interactions between individuals based on the structural features of the network [12]. Due to the bounded rationality of individuals, they are limited to obtaining information from adjacent nodes. Through pairwise or group interactions, diverse combinations of group strategies can be realized, potentially inducing evolutionary equilibrium states that unstructured groups cannot achieve [13].

Over the past two decades, researchers have used theoretical tools such as differential dynamics, probability theory, and stochastic processes, as well as computer simulation methods and real-world experiments, to derive typical conclusions on how cooperation strategies dominate in various types of networks. For instance, Nowak et al. [14] and Harut et al. [15] studied the evolution of network cooperation in structured regular networks through the prisoner's dilemma and snowdrift games, respectively. Their findings showed that while the payoff parameters differ, network structures can have both promoting and inhibiting effects on cooperative behavior, indicating that the impact of network structure on cooperation depends on the configuration of the payoff functions. In the context of a two-player, two-strategy game, Ohtsuki et al. [16] proposed a simple rule for the evolution of cooperation: cooperation will evolve if the benefit-to-cost ratio ($b/c$) of altruistic behavior exceeds the degree ($k$) of the network. However, this rule only approximately holds for various types of graphs. Allen et al. [17] introduced a universal mathematical model that generalizes this rule to populations on arbitrary graph structures, determining the critical benefit-cost ratio for the evolution of cooperation under weak selection conditions. From a biological perspective, Kuo et al. [18] derived an evolutionary dynamics model applicable to arbitrary complex spatial structures, which can be used to quantitatively predict the mutation fixation probabilities in large heterogeneous networks. Li et al. [19] extended research on static networks to dynamic temporal networks, demonstrating that dynamic interactions in social contexts are more conducive to the emergence and maintenance of cooperation than static networks. Civilini et al. [20] expanded pairwise games to higher-order networks, finding that higher-order interactions play a critical role in the survival of cooperation in social dilemmas. Alvarez-Rodriguez et al. [21] directly conducted public goods games on hypergraphs, identifying a trade-off between "benefit and cost" and determining the most favorable group size for cooperation. Sheng et al. [22] demonstrated that higher-order interactions significantly promote cooperation in multi-community connected networks. It is important to note that classical network game studies typically rely on two-player/multi-player, two-strategy (cooperate and defect) game models. The analysis of swarm intelligence in groups typically focuses on overall characteristics such as dominant strategies and absorption probabilities [23], but it is unable to determine the specific cooperation partners of individuals.

Previous research has improved traditional evolutionary games by adjusting interaction models, adding strategy options, or introducing coordination mechanisms, with the aim of achieving division of labor and cooperation among selfish individuals in a group. For example, Zhang et al. [24] introduced a "lazy" strategy to study the evolution of strategies in the context of self-organized division of labor dilemmas. Yaman et al. [25] proposed a decentralized social sanctioning mechanism that promotes the emergence of labor division by redistributing rewards within the group. However, these studies typically focus on three or four strategies, akin to cellular differentiation in biology, where the number of available strategies or tasks is far smaller than the number of evolving individuals. For distributed self-organizing networks with numerous tasks to be assigned, blindly expanding the strategy set increases the difficulty of theoretical analysis on complex networks. Furthermore, , characterizing individuals solely by their strategies fails to capture their cooperative relationships.

In the field of swarm intelligence, when studying cooperative behavior in structured populations, neighbors play a dual role: they serve as a limited source of information for strategy selection and also define the set of potential cooperation partners for each individual [26]. Therefore, directly adapting the evolutionary game models of unstructured finite populations to complex networks presents theoretical and practical limitations in both analysis and application.

The potential cooperation partners for a node include all its neighbors. From the perspective of traditional evolutionary game theory, each neighbor can be considered a potential cooperation partner (strategy). As a result, the strategy set in such a network would be equivalent to the set of nodes, leading to a high complexity in strategy evolution [27]. For networks with homogeneous node types, strategy evolution would struggle to converge, and the complexity of model analysis would be significantly elevated. Although traditional game-theoretic approaches can address these issues by modifying the strategy set [28] or introducing potential games [29], the universality of these solutions is correspondingly reduced. Therefore, there is an urgent need for an evolutionary game model on complex networks that can balance model simplicity while also accounting for the complexity of node strategies.

Driven by the aforementioned needs, this paper introduces an innovative game model—edge game—designed for the selection of cooperation partners in network cooperation evolution. In Section 2, we present the application scenarios and mathematical formulation of this model. Section 3 provides a theoretical analysis using regular networks as an example, deriving the conditions for cooperation evolution in the edge game. In Section 4, we conduct simulation experiments on complex networks with different levels of heterogeneity and compare the cooperative performance with traditional optimization algorithms. Finally, we conclude the paper with a summary and outlook.

## 2. Edge Game Model

### 2.1 Example Scenario

With the advancement of communication and unmanned aerial vehicle (UAV) technologies, the low-altitude economy is poised to become a significant force driving economic transformation. To illustrate the application scenario of the edge game, we use low-altitude logistics delivery as an example to describe the process of selecting cooperation partners in multi-agent collaboration.

As shown in **Fig. 1**(a), a non-fully connected transportation network is formed between four sender warehouses, 11 low-altitude UAVs, and five transfer hubs, influenced by factors such as spatial distance and communication capabilities. Each node can only access the material information from its adjacent nodes, with edges representing the potential for material transfer between nodes. In an ideal scenario, there are no constraints related to time, space, or resources, and every edge can be transformed into a cooperation edge.

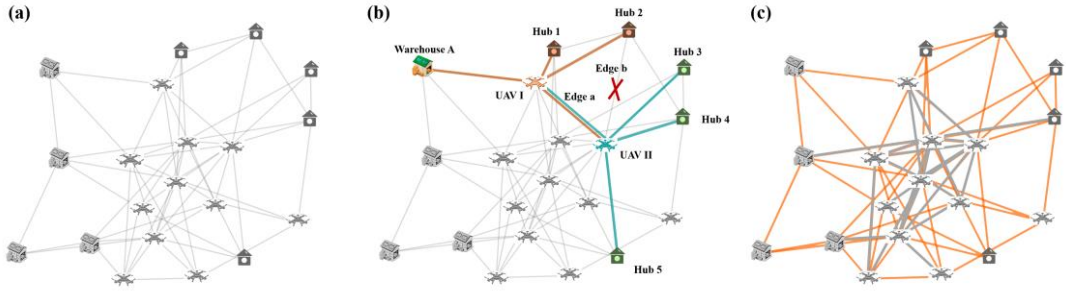

**Fig. 1** Schematic diagram of low-altitude logistics delivery network by UAV swarm. **a,** The material transportation network composed of dispatch warehouses, UAVs, and express hubs. **b,** Example of cooperative edges between nodes (UAV I and UAV II) under an unstable structure. **c,** The evolved stable cooperative network under a node cooperation threshold of 4, where orange edges represent cooperative edges and gray edges represent non-cooperative edges.

In the cooperative process with limited energy, factors such as scheduling time, drone battery life, and payload capacity restrict each node to selecting only a limited number of neighbors as cooperative partners in each batch of tasks. The node's payoff is positively correlated with the number of cooperation partners. Assuming each node can select at most 4 neighbors as cooperation partners, and the condition for successful cooperation is that the selected neighbors must also include the node itself as a partner, as illustrated by *Edge a* between UAV I and UAV II in **Fig. 1**(b). In this scenario, Warehouse A only requires UAV I and UAV II to transport goods to all the Express Hubs. However, this structure is not the optimal one. Taking Hub 2 as an example, its number of neighbors reaches the maximum threshold. Hub 2 could have established a cooperative relationship with UAV II, but since UAV II has already reached its limit of 4 cooperation partners, Hub 2's unidirectional cooperation demand cannot convert *Edge b* into a cooperation edge, leading to a suboptimal payoff for Hub 2. The cooperative structure will continue to evolve.

It is evident that the selection of cooperation edges is a dynamic and interactive process. Under external constraints and neighbor competition, an individual's strategy needs to evolve based on group interactions and fitness levels. Eventually, this process may lead to an evolutionary stable equilibrium (ESE), where no individual can improve their fitness by unilaterally changing their strategy. **Fig. 1**(c) depicts one such evolutionary stable equilibrium, where the orange edges represent cooperation edges, the gray edges represent non-cooperation edges, and each node has reached its maximum number of cooperation partners.

## 2.2 Role Reversal: The Introduction of Edge Game

When focusing on the "node" as the individual in an evolutionary game, the choice of which neighbors to select as cooperation partners becomes part of the node's strategy. In the scenario depicted in **Fig. 1**, the strategy set for each node will be highly variable and redundant. For instance, UAV II has 11 neighbors, so its strategy set will contain $2^{11}$ (2048) possible strategies to choose from. Moreover, since each node has a different set of neighbors, their strategy sets will vary, resulting in a complex and unpredictable evolutionary path that is difficult to converge.

In certain special graph theory models and algorithms, nodes and edges can "switch" roles under specific transformations. Whitney (1932) introduced the concept of line graphs (L(G)) by contracting the edges of a simple undirected graph (G) into nodes [30]. The edge set of G becomes the node set of L(G), and two nodes in L(G) are adjacent if and only if the corresponding edges in G are adjacent. It has been demonstrated that in optimization problems such as path selection and load balancing in networks, L(G) offers a more advantageous framework for analyzing interactions and dependencies among edges compared to G [31].

Inspired by the concept of line graphs, we propose using the "edge" as a hypothetical decision-making entity in an evolutionary game, which greatly simplifies the game model. In this case, edges have only two strategies. Each edge can acquire information from its adjacent edges and interact with them in the game. It is important to note that the goal of the edge game is to determine the cooperative edges of the "node" through the competition and cooperation among edges. Therefore, in this study, we do not strictly follow the line graph definition to group adjacent edges together. Instead, we divide the adjacent edges into two neighborhoods based on the two nodes connected by each edge (as shown in **Fig. 2**(b), where the blue adjacent edges are divided into two groups).

Similar to node-based games, edge games can also be divided into two types: pairwise interactions and group interactions. When considering the constraint on the number of cooperation partners, pairwise interactions are not conducive to calculating changes in the number of cooperative partners during evolution. Therefore, we opt for group interactions and propose an edge-based public goods game model by improving the traditional node-based public goods game model. The most significant difference between the proposed model and the traditional model lies in the number of game subgroups (see **Fig. 2**). In traditional node-based public goods games, the number of subgroups in which the focal node participates is $k+1$, whereas in edge games, regardless of how many adjacent edges a focal "individual" has, the number of game subgroups is always defined as 2.

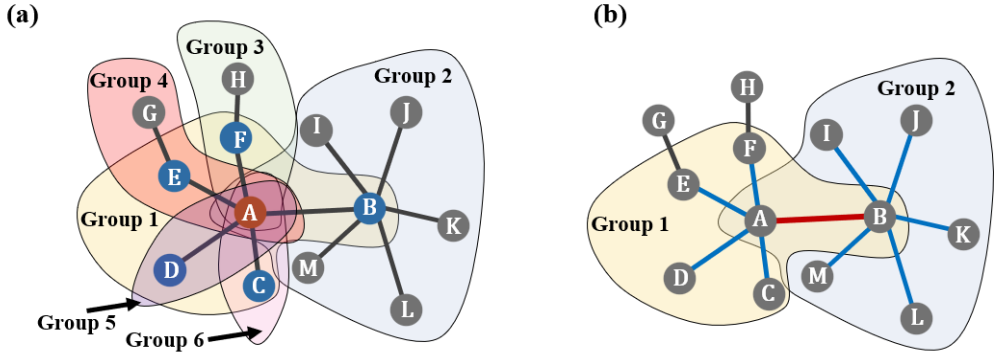

**Fig. 2** Illustrative diagram of group division in public goods games based on nodes (**a**) and edges (**b**), where the red node/edge represents the focal individual, and the blue nodes/edges represent the neighboring nodes/edges of the focal individuals.

### 2.3 Mathematical Definition of Edge Game

A group of size $N$ is connected by a simple undirected graph $G=(V,E)$, where $V=\{v_1,v_2,...,v_N\}$ represents the set of nodes, $k_n$ denotes the degree of node $v_n (n=1,2,...,N)$, and $E=\{e_1,e_2,...,e_M\}$ is the set of edges connecting the nodes. We conceptualize $e_m (m=1,2,...,M)$ as a game player, with nodes $v_p$ and $v_q$ as its endpoints, denoted as $e_m=\{v_p,v_q\}$. Each edge can obtain information about its neighboring edges and engage in a game interaction with its adjacent edges. The two groups that $e_m$ involved in are denoted as $\Omega_p$ and $\Omega_q$, where $\Omega_p=\{\{v_p,v_i\}|v_i \in V,\{v_p,v_i\}\in E\}$ is the set of edges directly connected to node $v_p$, and $\Omega_q=\{\{v_q,v_j\}|v_j \in V,\{v_q,v_j\}\in E\}$ is the set of edges directly connected to node $v_q$. The number of elements in these two sets is $k_p$ and $k_q$, respectively.

Define the population strategy set $S = \{C, D\}$, where $C$ stands for the cooperators and $D$ represents the defectors. Each game player $e_m \in E$ selects a strategy $s_{e_m} \in S$. In each game group, a cooperator pays a *cost* as a contribution to the public good, which generates a return *r·cost* for the group, where $r > 1$ is the synergy factor, indicating that the return exceeds the contribution. In contrast, a defector pays no cost but enjoys the benefits generated by the contributions of others.

### 2.3.1 Payoff Function

Let the number of cooperators in the groups $\Omega_p$ and $\Omega_q$ be denoted as $n_p$ and $n_q$, respectively. When the maximum cooperative neighbor threshold of nodes is not considered, the payoff of an edge is influenced only by its strategy, the number of elements in the two game groups, and the number of cooperators, given the *r* and *cost*.

If $s_{e_m} = C$, then the payoffs of $e_m$ in the two game groups are as follows:

$$\pi_{\Omega_p}^{C}(n_p) = \frac{n_p \cdot r \cdot cost}{k_p} - cost \tag{1}$$

$$\pi_{\Omega_q}^{C}(n_q) = \frac{n_q \cdot r \cdot cost}{k_q} - cost \tag{2}$$

The total payoff of $e_m$ is:

$$\Pi_{e_m}^{C}(n_p, n_q) = \pi_{\Omega_p}^{C}(n_p) + \pi_{\Omega_q}^{C}(n_q) = \frac{n_p \cdot r \cdot cost}{k_p} + \frac{n_q \cdot r \cdot cost}{k_q} - 2 \cdot cost \tag{3}$$

If $s_{e_m} = D$, then the payoffs of $e_m$ in the two game groups are as follows:

$$\pi_{\Omega_p}^{D}(n_p) = \frac{n_p \cdot r \cdot cost}{k_p} \tag{4}$$

$$\pi_{\Omega_q}^{D}(n_q) = \frac{n_q \cdot r \cdot cost}{k_q} \tag{5}$$

The total payoff of $e_m$ is:

$$\Pi_{e_m}^{D}(n_p, n_q) = \pi_{\Omega_p}^{D}(n_p) + \pi_{\Omega_q}^{D}(n_q) = \frac{n_p \cdot r \cdot cost}{k_p} + \frac{n_q \cdot r \cdot cost}{k_q} \tag{6}$$

When considering the maximum cooperative neighbor threshold constraint for the nodes, the payoff of an edge is influenced not only by the edge strategy, the number of elements in the two groups, and the number of cooperators but also by the node's maximum cooperative neighbor threshold ($\theta \geq 1$). Here, we stipulate that in each game group, when the number of cooperators does not exceed $\theta$, the payoffs remain the same as previously described. However, if the number of cooperators exceeds *θ*, no members of the group will receive a reward. In this case, the payoff of defectors is 0, and cooperators must additionally pay the cost, resulting in a payoff of -*cost*.

**2.3.2 Strategy Update**

In edge games, we employ the direct protocol for strategy updating, which differs from the imitation protocol. Unlike imitation protocols where individuals copy the strategies of their neighbors, each individual in the direct protocol is aware of the complete strategy set and can directly select a candidate strategy without the need to imitate an existing neighbor's strategy [32]. The direct protocol is more suitable for strategy updating in cooperation processes that consider thresholds, as individuals can directly compare the payoffs of new and old strategies in the group game to decide whether to update their strategy.

Under the assumption of the direct protocol, a typical framework is the Smith dynamics equation proposed by Smith in his study of traffic flow dynamics [33]. Here, we adapt this evolutionary process model and use pairwise comparison to evaluate the payoffs of current and new strategies. If the payoff of the new strategy *j* is higher than that of the current strategy *i*, the individual will switch to the new strategy. The transition rate from strategy *i* to strategy *j* for the individual $e_m$ is given by:

$$\tau_{i \to j}(e_m) = \left[ \Pi^j_{e_m} - \Pi^i_{e_m} \right]_+, \forall i, j \in S, \forall e_m \in E \tag{7}$$

where $[y]_+ = max(y,0)$. It is evident that $\tau_{i \to j}(e_m) \geq 0, \forall e_m \in E$.

By averaging the transition rates (*τ*) across the population, we derive the population's state adjustment protocol ( T ). Substituting this into the framework yields the Smith dynamics equation for the population state:

$$\dot{x}_i = \sum_{j \in S} x_j T_{j \to i} - x_i \sum_{j \in S} T_{i \to j}, \forall i \in S \tag{8}$$

where $x_i$ represents the proportion of individuals adopting strategy *i*. The change in the proportion of individuals with strategy *i*, $\dot{x}_i$ , is determined by the inflow of individuals switching from other strategies to strategy *i*, minus the outflow of individuals switching from strategy *i* to other strategies.

## 3. Theoretical Analysis

Based on the above mathematical model, we use a special type of nearest-neighbor coupled (NC) network as an example to derive the conditions for cooperation evolution in edge games on regular networks. Assume that there are $N$ nodes in the NC network, each node is connected to $k$ ($k$ is even) neighbors, and the number of edges in the network is $M = N \cdot k / 2$.

### 3.1 Conditions for Cooperation Evolution without Threshold Constraints

Assume that all nodes have sufficient information processing capabilities and energy to cooperate with all their neighbors, i.e., without considering the threshold constraints on the number of cooperative neighbors for each node, we continue the traditional study of cooperation evolution in network evolutionary games by first exploring the conditions under which all edges in the network become cooperative edges.

Taking the classic two-strategy game as an example, where only two strategy transitions are possible, from **Eq. (8)** we obtain:

$$\begin{cases} \dot{x}_C = x_D \mathrm{T}_{D \to C} - x_C \mathrm{T}_{C \to D} \\ \dot{x}_D = x_C \mathrm{T}_{C \to D} - x_D \mathrm{T}_{D \to C} \\ \quad x_D + x_C = 1 \end{cases} \tag{9}$$

Simplifying, we get:

$$\dot{x}_C = (1 - x_C) \mathrm{T}_{D \to C} - x_C \mathrm{T}_{C \to D} \tag{10}$$

In the NC network, $k_p = k_q = k$. **Eqs. (3)** and **(6)** can be simplified to:

$$\Pi^C_{e_m}(n_p, n_q) = \frac{(n_p + n_q) \cdot r - 2k}{k} \cdot cost \tag{11}$$

$$\Pi^D_{e_m}(n_p, n_q) = \frac{(n_p + n_q) \cdot r}{k} \cdot cost \tag{12}$$

From **Eq. (7)**, we can derive:

$$\begin{aligned} \tau_{D \to C}(e_m) &= \left[ \Pi^C_{e_m}(n_p + 1, n_q + 1) - \Pi^D_{e_m}(n_p, n_q) \right]_+ \\ &= [\frac{2(r-k)}{k} \cdot cost]_+, \forall e_m \in E, n_p < k, n_q < k \end{aligned} \tag{13}$$

$$\tau_{C \to D}(e_m) = \left[ \Pi_{e_m}^D(n_p - 1, n_q - 1) - \Pi_{e_m}^C(n_p, n_q) \right]_+$$
$$= [\frac{2(k-r)}{k} \cdot cost]_+, \forall e_m \in E, n_p > 0, n_q > 0 \quad (14)$$

Discuss the relative magnitude of $r$ and $k$. For $\forall e_m \in E$, we have:

$$\tau_{D \to C}(e_m) = \begin{cases} 0 & ,r \leq k \\ \dfrac{2(r-k)}{k} \cdot cost & ,r > k \end{cases} \quad (15)$$

$$\tau_{C \to D}(e_m) = \begin{cases} \dfrac{2(k-r)}{k} \cdot cost & ,r < k \\ 0 & ,r \geq k \end{cases} \quad (16)$$

It is evident that the strategy transition rate in the NC network is independent of the position of the edges in the network. Thus, the average state is obtained as:

$$\forall e_m \in E, \quad \begin{aligned} T_{D \to C} &= \tau_{D \to C}(e_m) \\ T_{C \to D} &= \tau_{C \to D}(e_m) \end{aligned} \quad (17)$$

Substituting into **Eq.** (10), we get the time evolution of the proportion of cooperators in the population as:

$$\dot{x}_C = \begin{cases} -x_C \cdot \dfrac{2(k-r)}{k} \cdot cost & ,r < k \\ 0 & ,r = k \\ (1-x_C) \dfrac{2(r-k)}{k} \cdot cost & ,r > k \end{cases} \quad (18)$$

To solve for $x_C$ when $r > k$, let $\eta = \dfrac{2(r-k)}{k} \cdot cost$ and the initial value of $x_C$ be $x_0$. It can be shown that $\eta > 0$ for $x_0 \in [0,1]$. The detailed derivation is provided in Appendix A, yielding:

$$x_C = 1 - (1-x_0)e^{-\eta t} = 1 - (1-x_0)e^{-\frac{2(r-k)cost}{k}t} \quad (19)$$

It is evident that when $r > k$, for $\forall x_0 \in [0,1]$, $x_C \to 1$ as $t \to \infty$, indicating the network evolves to a state where all agents cooperate.

Similarly, when $r < k$, $x_C = x_0 e^{-\frac{2(k-r)cost}{k}t}$, for $\forall x_0 \in [0,1]$, $x_C \to 0$ as $t \to \infty$, and the system tends towards all defectors; when $r < k$, $x_C = x_0$, for $\forall x_0 \in [0,1]$, $x_C = x_0$ as $t \to \infty$, and the proportion of cooperators stabilizes.

Therefore, the condition for cooperation evolution in the NC network without threshold constraints is $r > k$.

## 3.2 Conditions for Cooperation Evolution under Threshold Constraints

When considering the maximum cooperation neighbor threshold of each node, given $1 \leq \theta \leq k$, we need to discuss the relative sizes of $n_p$, $n_q$ and $\theta$ according to the definition in Section 2.3.1. For $\forall e_m \in E$, **Eqs.** (11) and (12) are modified to:

$$\Pi_{e_m}^C(n_p, n_q) = \begin{cases} \dfrac{(n_p + n_q) \cdot r - 2k}{k} \cdot cost & n_p \leq \theta, n_q \leq \theta \\ \dfrac{n_p \cdot r - 2k}{k} \cdot cost & n_p \leq \theta, n_q > \theta \\ \dfrac{n_q \cdot r - 2k}{k} \cdot cost & n_p > \theta, n_q \leq \theta \\ -2cost & n_p > \theta, n_q > \theta \end{cases} \quad (20)$$

$$\Pi_{e_m}^D(n_p, n_q) = \begin{cases} \dfrac{(n_p + n_q) \cdot r}{k} \cdot cost & n_p \leq \theta, n_q \leq \theta \\ \dfrac{n_p \cdot r}{k} \cdot cost & n_p \leq \theta, n_q > \theta \\ \dfrac{n_q \cdot r}{k} \cdot cost & n_p > \theta, n_q \leq \theta \\ 0 & n_p > \theta, n_q > \theta \end{cases} \quad (21)$$

Since $\tau_{D \to C}(e_m) = \left[ \Pi_{e_m}^C(n_p+1, n_q+1) - \Pi_{e_m}^D(n_p, n_q) \right]_+$, $\tau_{C \to D}(e_m) = \left[ \Pi_{e_m}^D(n_p-1, n_q-1) - \Pi_{e_m}^C(n_p, n_q) \right]_+$, we discuss the relative sizes of $n_p$, $n_q$, $n_p+1$, $n_q+1$, $n_p-1$, $n_q-1$ and $\theta$ to derive the expressions for $\tau_{D \to C}(e_m)$ and $\tau_{C \to D}(e_m)$ as shown in **Table 1**.

**Table 1** Expressions for $\tau_{D \to C}(e_m)$ and $\tau_{C \to D}(e_m)$ under Threshold Constraints

| | $n_p = 0$ | $0 < n_p \leq \theta - 1$ | $n_p = \theta$ | $n_p = \theta + 1 < k$ | $\theta + 1 < n_p < k$ | $n_p = k$ |
|---|---|---|---|---|---|---|
| | | | $\tau_{D \to C}(e_m)$ | | | |
| $n_q = 0$ | $[\frac{2(r-k)}{k} \cdot cost]_+$ | $[\frac{2(r-k)}{k} \cdot cost]_+$ | $[\frac{r(1-\theta)-2k}{k} \cdot cost]_+$ | $[\frac{2(r/2-k)}{k} \cdot cost]_+$ | $[\frac{2(r/2-k)}{k} \cdot cost]_+$ | / |
| $0 < n_q \leq \theta - 1$ | $[\frac{2(r-k)}{k} \cdot cost]_+$ | $[\frac{2(r-k)}{k} \cdot cost]_+$ | $[\frac{r(1-\theta)-2k}{k} \cdot cost]_+$ | $[\frac{2(r/2-k)}{k} \cdot cost]_+$ | $[\frac{2(r/2-k)}{k} \cdot cost]_+$ | 0 |
| $n_q = \theta$ | $[\frac{r(1-\theta)-2k}{k} \cdot cost]_+$ | $[\frac{r(1-\theta)-2k}{k} \cdot cost]_+$ | $[\frac{2(r\theta - k)}{k} \cdot cost]_+$ | $[\frac{2(-r\theta/2-k)}{k} \cdot cost]_+$ | $[\frac{2(-r\theta/2-k)}{k} \cdot cost]_+$ | 0 |
| $n_q = \theta + 1 < k$ | $[\frac{2(r/2-k)}{k} \cdot cost]_+$ | $[\frac{2(r/2-k)}{k} \cdot cost]_+$ | $[\frac{2(-r\theta/2-k)}{k} \cdot cost]_+$ | $[-2cost]_+$ | $[-2cost]_+$ | 0 |
| $\theta + 1 < n_q < k$ | $[\frac{2(r/2-k)}{k} \cdot cost]_+$ | $[\frac{2(r/2-k)}{k} \cdot cost]_+$ | $[\frac{2(-r\theta/2-k)}{k} \cdot cost]_+$ | $[-2cost]_+$ | $[-2cost]_+$ | 0 |
| $n_q = k$ | / | 0 | 0 | 0 | 0 | 0 |
| | | | $\tau_{C \to D}(e_m)$ | | | |
| | $n_p = 0$ | $0 < n_p \leq \theta - 1$ | $n_p = \theta$ | $n_p = \theta + 1 < k$ | $\theta + 1 < n_p < k$ | $n_p = k$ |

| $n_q$ | | | | | | |
|---|---|---|---|---|---|---|
| $n_q = 0$ | 0 | 0 | 0 | 0 | 0 | / |
| $0 < n_q \leq \theta - 1$ | 0 | $[\frac{2(k-r)}{k} \cdot cost]_+$ | $[\frac{2(k-r)}{k} \cdot cost]_+$ | $[\frac{2k-r(1-\theta)}{k} \cdot cost]_+$ (dashed) | $[\frac{2(k-r/2)}{k} \cdot cost]_+$ | $[\frac{2(k-r/2)}{k} \cdot cost]_+$ |
| $n_q = \theta$ | 0 | $[\frac{2(k-r)}{k} \cdot cost]_+$ | $[\frac{2(k-r)}{k} \cdot cost]_+$ | $[\frac{2k-r(1-\theta)}{k} \cdot cost]_+$ (dashed) | $[\frac{2(k-r/2)}{k} \cdot cost]_+$ | $[\frac{2(k-r/2)}{k} \cdot cost]_+$ |
| $n_q = \theta + 1 < k$ | 0 | $[\frac{2k-r(1-\theta)}{k} \cdot cost]_+$ (dashed) | $[\frac{2k-r(1-\theta)}{k} \cdot cost]_+$ (dashed) | $[\frac{2(r\theta+k)}{k} \cdot cost]_+$ | $[\frac{2(r\theta/2+k)}{k} \cdot cost]_+$ (dashed) | $[\frac{2(r\theta/2+k)}{k} \cdot cost]_+$ (dashed) |
| $\theta + 1 < n_q < k$ | 0 | $[\frac{2(k-r/2)}{k} \cdot cost]_+$ | $[\frac{2(k-r/2)}{k} \cdot cost]_+$ | $[\frac{2(r\theta/2+k)}{k} \cdot cost]_+$ (dashed) | $[2cost]_+$ (dashed) | $[2cost]_+$ (dashed) |
| $n_q = k$ | / | $[\frac{2(k-r/2)}{k} \cdot cost]_+$ | $[\frac{2(k-r/2)}{k} \cdot cost]_+$ | $[\frac{2(r\theta/2+k)}{k} \cdot cost]_+$ (dashed) | $[2cost]_+$ (dashed) | $[2cost]_+$ (dashed) |

Clearly, when $n_p = 0$ or $n_q = 0$, $\tau_{C \to D}(e_m)$ is always 0. Similarly, when $n_p = k$ or $n_q = k$, $\tau_{D \to C}(e_m)$ is always 0. Since the focal edge is shared by two groups, the combination (0, $k$) for $n_p$ and $n_q$ does not exist. When $(n_p \in (0,k)) \wedge (n_q \in (0,k))$, it requires discussion based on different circumstances: based on the initial condition $1 \leq \theta \leq k, r > 1$, the items underlined with solid lines in **Table 1** can be directly simplified to 0, while those underlined with dashed lines are always positive. The remaining terms need to be discussed based on the relationship between $r$, $k$ and $\theta$. The results for four cases — $1 < r \leq k/\theta$, $k/\theta < r \leq k$, $k < r \leq 2k$, and $r > 2k$ — are detailed in Appendix B.

From Appendix B, we can conclude that when $1 < r \leq k/\theta$, only $\tau_{D \to C}(e_m)$ is a constant (0) unaffected by the network structure ($n_p$ and $n_q$), while under other conditions, $\tau_{D \to C}(e_m)$ and $\tau_{C \to D}(e_m)$ exhibit piecewise variations depending on the network structure. Therefore, unlike the case without threshold constraints, we cannot directly derive the average state T (Eq. (17)) for the Smith dynamics equation. Consequently, we cannot obtain a macroscopic condition for cooperation.

Here, we discuss the cooperation between nodes from a micro perspective, using auxiliary flow diagrams to derive the conditions for cooperation and the proportion of cooperative edges. Consider an NC network with 100 nodes and 8 neighbors per node. We analyze the strategy transition rates of the focal edge ($e_m$) under different cooperation thresholds. As shown in **Fig. 3** and **Fig. 4**, the horizontal and vertical coordinates represent the number of cooperative edges in the two game groups of $e_m$. The grid in the figure depicts all combinations of cooperative edge numbers in the groups that $e_m$ participates. The grid points within the white box have two possible strategy transitions (D→C and C→D), which are of interest to us. The relative sizes of $\tau_{D \to C}(e_m)$ and $\tau_{C \to D}(e_m)$ at each grid point are represented by the lengths of white and yellow arrows, respectively. The white arrow points to the number of cooperators in the groups after the focal edge switches from defection to cooperation, while the yellow arrows indicate the reverse. The difference between $\tau_{D \to C}(e_m)$ and $\tau_{C \to D}(e_m)$ at each grid point is denoted as $\delta$: if $\tau_{D \to C}(e_m) > \tau_{C \to D}(e_m)$, then $\delta > 0$, and the grid is marked in red, indicating that the edge is more likely to switch from defection to cooperation under this network structure; if $\tau_{D \to C}(e_m) < \tau_{C \to D}(e_m)$, then $\delta < 0$, and the grid is marked in blue, indicating that the edge is more likely to switch from cooperation to defection; if $\tau_{D \to C}(e_m) = \tau_{C \to D}(e_m)$, then $\delta = 0$, and the grid is marked in gray, indicating that the edge is more likely to maintain its current strategy. The values at $(0, k)$ and $(k, 0)$ are NaN and are also marked in gray.

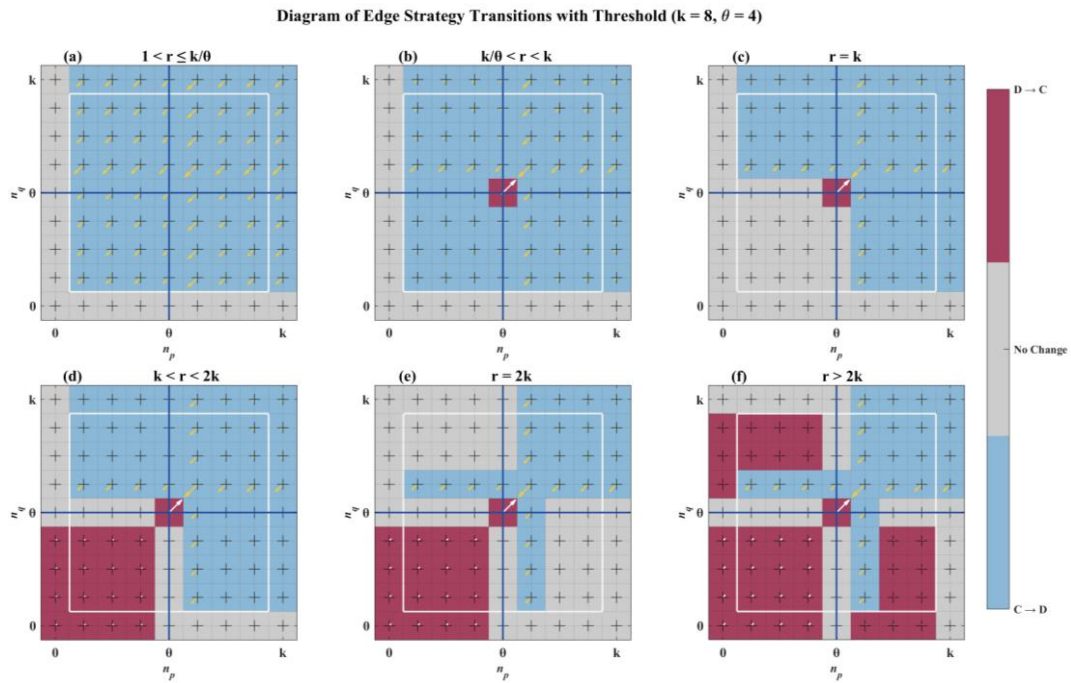

**Fig. 3** Distribution patterns of strategy transition rates due to changes in the value of $r$ under threshold conditions (NC network with $k = 8$, $\theta = 4$). (**a**) to (**f**) show six distribution modes as $r$ increases. The horizontal and vertical coordinates represent the number of cooperators in the two game groups of the focal edge. G rid points within the white box have two possible strategy transitions. Red grids indicate that the edge is more likely to switch from defection to cooperation, blue grids indicate that the edge is more likely to switch from cooperation to defection, and gray grids indicate that the edge is more likely to maintain its original strategy. White arrows point to the number of cooperators in the group after the focal edge switches from defection to cooperation, while yellow arrows point to the number of cooperators after the focal edge switches from cooperation to defection. The dark blue vertical and horizontal lines represent $n_p = \theta$ and $n_q = \theta$, respectively. The same applies to the following figure.

First, we fix the maximum number of cooperative neighbors ($\theta$) at 4 and compare the strategy transition rate differences ($\delta$) under different $r$ values, obtaining six distribution patterns as shown in **Fig. 3**. It can be seen that:

(1) When $1 < r \leq k/\theta$ (**Fig. 3**(a)), except for $\delta = 0$ when $n_p = 0$ or $n_q = 0$ (where only $s_{e_m} = D$ exists), all other grids have $\delta < 0$, indicating that all edges with a cooperation strategy will switch to defection until all edges in the network adopt defection. Thus, the macroscopic ESS is all defectors.

(2) When $k/\theta < r < k$, compared to **Fig. 3**(a), only the grid point at $(n_p, n_q) = (\theta, \theta)$ shows a D→C trend in **Fig. 3**(b). However, due to the stochastic evolution process, once the number of cooperative edges in the group of the focal edge fluctuates, the state of $n_p = \theta$ and $n_q = \theta$ will be disrupted. Therefore, the grid point with $\delta > 0$ is an unstable point, and the ESS in **Fig. 3**(b) is the same as in **Fig. 3**(a).

(3) When $r = k$ (**Fig. 3**(c)), when $n_p > \theta$ or $n_q > \theta$, $\delta$ is similar to **Fig. 3**(a) and **Fig. 3**(b), mostly switching from defection to cooperation until the number of cooperators on both sides is less than or equal to $\theta$. Once this state is reached, the strategy of the focal edge remains unchanged. It is worth noting that when $n_p < \theta$ and $n_q < \theta$, the strategy of the focal edge will remain unchanged. Therefore, the ESS in **Fig. 3**(c) is affected by the initial number of cooperative edges, and the final cooperation proportion is a variable within the interval $[0, \theta/k)$.

(4) When $r > k$, the region where edge strategies switch to cooperation begins to increase. In **Fig. 3**(d) ($k < r < 2k$), the stable points are located in the two gray zigzags separated by the grid of $(\theta, \theta)$. Outside the white box, the strategies of the focal edges are defection, while inside the box, the number of cooperative edges on both sides of the focal edge will tend towards $\theta/k$ but remain less than $\theta/k$. Considering the stable regions inside and outside the box, the proportion of cooperative edges is less than $\theta/k$. The stable points in **Fig. 3**(e) extend to the regions where $n_p > \theta$ or $n_q > \theta$, and the number of cooperators exceeding the threshold begins to appear. Affected by the initial number of cooperators, if the initial number of cooperators is located in the upper-left or lower-right stable regions, the cooperation proportion is likely to be greater than $\theta/k$. If the initial number of cooperators is small and located in the lower-left region, the cooperation proportion is likely to be less than $\theta/k$. When $r$ is sufficiently large (**Fig. 3**(f), $r > 2k$), only the stable points of the focal edges with $n_p \leq \theta+1$ and $n_q \leq \theta+1$ will evolve to the reference line where the number of cooperators in the group is $\leq \theta$. In other cases, the stable points of the focal edges will evolve to the stable region where the number of cooperators in the group is $\geq \theta$. Therefore, affected by the initial distribution, when the initial number of cooperators is very small, the cooperation proportion will remain below $\theta/k$ for $r > 2k$. When the initial number of cooperators increases and breaks through the threshold limit, the number of cooperators exceeding the threshold will also increase as in **Fig. 3**(e).

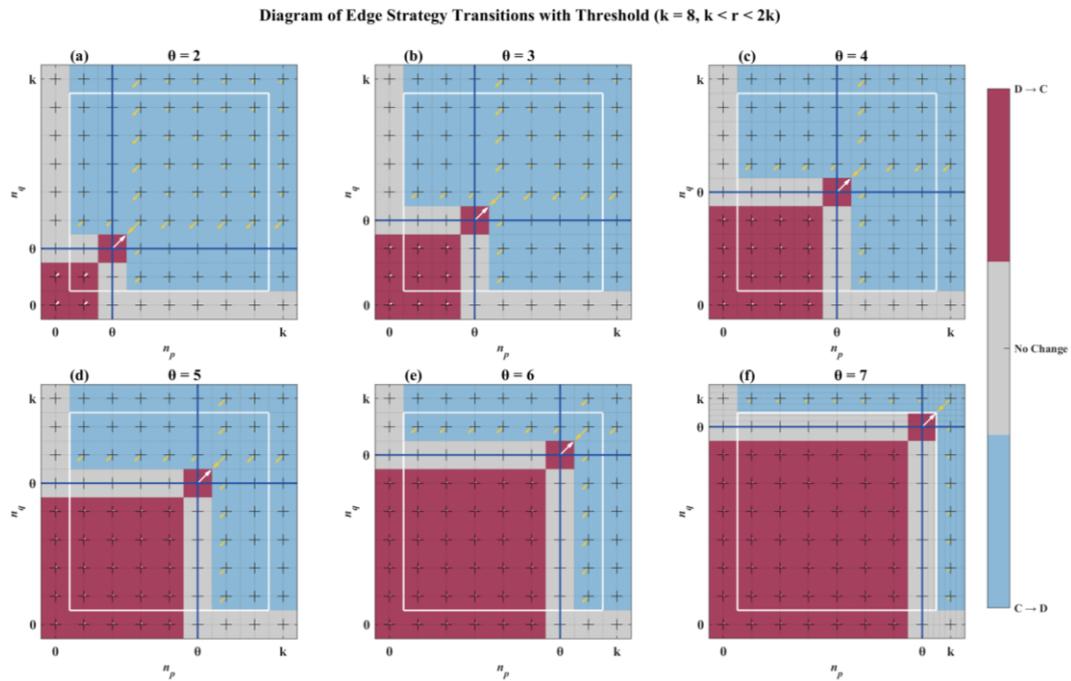

**Fig. 4** Distribution patterns of strategy transition rates due to changes in the value of $\theta$ under threshold conditions (NC network with $k = 8$, $k < r < 2k$).

The maximum cooperation neighbor threshold $\theta$ is related to the node's load performance. In **Fig. 4**, using an NC network with $k = 8$ and $k < r < 2k$, we analyze the effect of varying $\theta$ on the stable state. The results indicate that changes in $\theta$ do not affect the pattern of stable points, which remain within the two gray zigzags. The final proportion of cooperative edges is a small value less than $\theta/k$. Similar results are observed when varying the average degree $k$ and network size $N$ of the NC network (figure omitted), which indicate that these two variables are not key influencing factors.

## 3.3 Calculation of Average Population Payoff

Section 3.1 solves the average payoff of the population after evolution stabilization without threshold constraints based on **Eqs.** (11) and (12):

When the ESS is C, we have $n_p = n_q = k$, and the payoff is:

$$\overline{\Pi}_{e_m} = \frac{2k \cdot r - 2k}{k} \cdot cost = 2(r-1) \cdot cost \tag{22}$$

When the ESS is D, we have $n_p = n_q = 0$, and the payoff is: $\overline{\Pi}_{e_m} = 0$.

It is evident that without threshold constraints, the optimal payoff of the population after evolutionary stability (Eq. (22)) is easy to determine. However, as discussed in Section 3.2, when threshold constraints are considered, it is challenging to provide a single expression for the population's payoff.

Here, we start by examining the payoff of the focal edge $e_m$ and speculate on the maximum possible payoff for the population. Based on the definitions of **Eqs.** (20) and (21), we observe that $r$ and *cost* have a linear effect on individual payoff, which does not affect the trend or structural changes. Therefore, we consider an NC network with $k = 8$ and $\theta = 4$, and define $r = 12$ (where $k < r < 2k$) and $cost = 1$. The payoff of the focal edge, when adopting either a cooperation or defection strategy, as a function of the number of cooperating neighbors, is shown in **Fig. 5**. Regardless of the strategy chosen by the edge, the payoff peaks when the number of cooperators in both groups is equal to $\theta$, and the peak payoffs for both strategies are denoted as $\Pi^C_{e_m}(\theta,\theta) = (2\theta r - 2k) \cdot cost / k = 10$ and $\Pi^D_{e_m}(\theta,\theta) = 2\theta r \cdot cost / k = 12$. Further experiments reveal that adjusting $\theta$ (where $1 \leq \theta \leq k$) and $k$ still results in peak values at the point $(\theta, \theta)$ (figure omitted).

In theory, the number of cooperative edges in an NC network can reach the scenario where $(n_p, n_q) = (\theta, \theta)$. For example, in an NC network with $N = 100$, $k = 8$, $\theta = 4$, this situation is depicted in **Fig. 6**(a). When $(n_p, n_q) = (4, 4)$, the cooperative edges are illustrated in **Fig. 6**(b). In this case, both game groups formed by these edges each have exactly 4 cooperators. The proportion of cooperators at this point is $\theta/k = 0.5$. Ideally, with half cooperators and half defectors in the group, the maximum average payoff for the group is $max(\overline{\Pi}_{e_m}) = (2\theta r - k) \cdot cost / k$. In the current example, $max(\overline{\Pi}_{e_m}) = 11$.

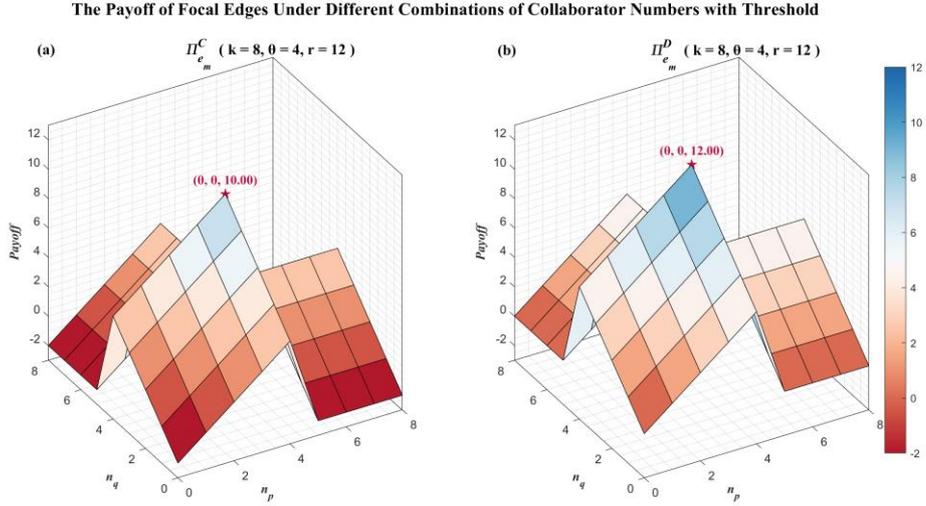

**Fig. 5** Under threshold constraints, the payoff of $e_m$ varies with the number of cooperating neighbors when holding the cooperation (**a**) or defection (**b**) strategy (using the NC network with $k = 8$, $\theta = 4$ and $k < r < 2k$ as an example; similar trends are observed for other values of $r$).

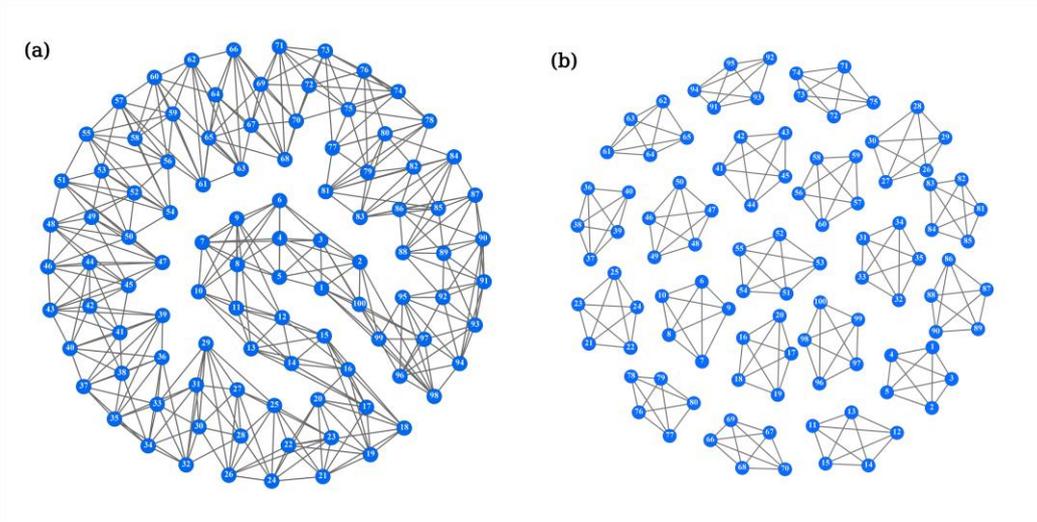

**Fig. 6 a**, Schematic of the NC network with $N = 100$, $k = 8$; **b**, The cooperative structure that achieves the maximum payoff when the threshold $\theta = 4$ in the NC network.

However, as discussed in **Fig. 3**, regardless of the value of $r$, the scenario $(n_p, n_q) = (\theta, \theta)$ is not a stable point in the evolutionary process. The stable state of the population will not remain in this ideal state. Therefore, under threshold constraints, we have:

$$\overline{\Pi}_{e_m} < (2\theta r - k) \cdot cost / k \tag{23}$$

In the current example, this results in a payoff of $\overline{\Pi}_{e_m} < 11$.

### 3.4 Moderate Cooperation and Excessive Cooperation

Although the maximum payoff at $(n_p, n_q) = (\theta, \theta)$ shown in **Fig. 3** is not a stable solution under various game conditions, as indicated in **Fig. 5**, it can be observed that in the vicinity of $(\theta, \theta)$, individual payoffs increase linearly when $np, nq < \theta$ until reaching the point $(\theta, \theta)$. Once $n_p, n_q > \theta$, individual payoffs experience a sharp drop. Therefore, values of $n_p < \theta$ or $n_q < \theta$ are more likely to reach the optimal solution of the game.

Thus, the stable points located in the regions where $n_p > \theta$ or $n_q > \theta$ in **Fig. 3**(e) and **Fig. 3**(f) represent stable solutions under the current game conditions, but from the perspective of individual payoffs, they do not correspond to the best outcomes for individuals. In this context, we define the phenomenon of excessively increasing the proportion of cooperators (in NC networks, this may exceed $\theta/k$) and having the number of cooperators at nodes exceed the threshold constraint as "excessive cooperation", which corresponds to the real-world phenomenon of "High stakes draw out the daring." In such a scenario, not only are the stable solutions for each edge not optimal, but this strategy distribution may also lead to node overload or collapse, which is not the best cooperative structure.

In contrast, the cooperative phenomena below the threshold, shown in **Fig. 3**(d), are more reasonable, where the number of cooperative edges at each node is still within the threshold. We define such cooperation, where the proportion of cooperators is approximately optimal (less than $\theta/k$) and the number of cooperators at all nodes remains below the threshold, as "moderate cooperation".

To achieve "moderate cooperation" in group cooperation, given a cooperation threshold, it can be realized by adjusting the initial number of cooperators and the magnitude of the synergy factor $r$.

### 4. Experiments and Discussion

In Section 3, we derived the critical conditions for achieving effective cooperation with and without threshold constraints using the NC network as an example. We identified the key role of the synergy factor $r$ in the proportion of cooperators in the population and defined and described the scenarios of "excessive cooperation" and "moderate cooperation" in relation to the relative size of the average population payoff. In the following, we will validate the theoretical conclusions derived for the NC network through simulation experiments and explore the conditions and cooperation rates of cooperation evolution in other types of networks.

### 4.1 Evolutionarily Stable Strategy in NC Networks

To verify the conclusions drawn from the theoretical analysis of regular networks, the simulation experiments were conducted using an NC network with parameters $N = 100$, $k = 8$, $M = 400$ as the baseline. As discussed earlier, under fixed payment costs, the *cost* is simply a linear parameter and does not affect the trend or structure of the cooperators' proportion (**Eqs.** (15) and (16),**Table 1**). For the sake of discussion, we set *cost* = 1 in the following sections.

#### 4.1.1 Cooperation Evolution Conditions without Threshold Constraints

To explore the effect of the parameter $r$ on cooperation evolution, the range of $r$ was set to $(1, 2k]$, with a step size of 0.1, to capture the characteristics of the proportion of cooperators and the average population payoff as $r$ changes. The initial number of cooperators was set to 0, and all other parameters were fixed to baseline values to isolate the independent effect of $r$. The criterion for evolution to reach a stable state was that the number of cooperators did not change for 500 consecutive iterations.

The simulation results for ESS are shown in **Fig. 7**. The blue line corresponds to the proportion of cooperators ($y_1$) on the left y-axis, and the orange line corresponds to the average population payoff ($y_2$) on the right y-axis. Both variables exhibit a significant step-change characteristic as $r$ increases: when $r \leq k$, both $y_1$ and $y_2$ are 0; at $r = k$, a jump occurs; when $r > k$, both $y_1$ and $y_2$ increase in a stepwise manner. At $r = 8.1$, $y_1 = 1$, $y_2 = 14.2$, after which $y_1$ remains at 1, and $y_2$ increases linearly, with the straight-line coinciding with the function curve of $y_2 = 2r - 2$. The payoff values are perfectly consistent with the theoretical analysis in **Eq.** (22).

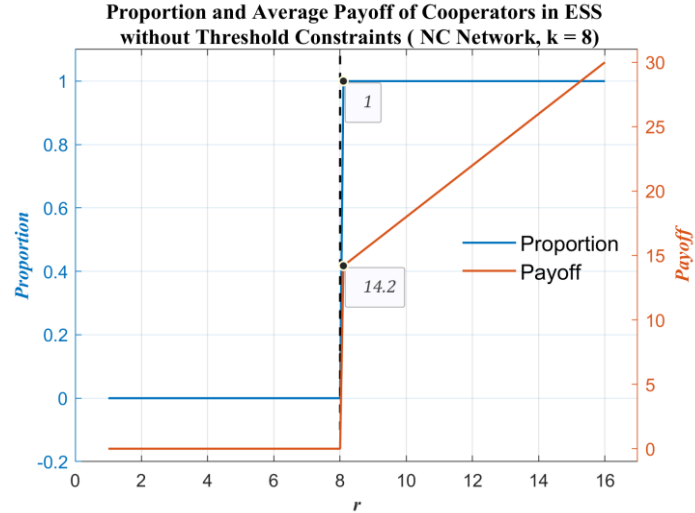

**Fig. 7** Proportion of cooperators (blue line) and average group payoff (orange line) in NC networks at ESS without threshold constraints for different $r$ values. The black dashed line as a reference at $r = k$.

Further investigation was conducted on the impact of the initial proportion of cooperators ($x_0$) on the steady state. Since the initial number of cooperators can be any integer between 0 and 400, $x_0$ was sampled 401 times at equal intervals between 0 and 1 for comparison (**Fig. 8** and **Fig. 9**). For illustration, the values of $r$ were taken as $(1+k)/2 = 4.5$, $k = 8$, and $1.5k = 12$, representing three cases: $1 < r < k$ (blue), $r = k$ (orange), and $r > k$ (purple). The horizontal axis represents the number of iterations in the evolution, and **Fig. 8** and **Fig. 9** summarize the evolution of the cooperators' proportion and average payoff over 401 experimental runs. It is evident that regardless of the value of $x_0$, after the evolution stabilizes, when $1 < r < k$, the proportion of cooperators is 0, and the average population payoff is 0; when $r > k$, the proportion of cooperators is 1, and the average population payoff is 22; when $r = k$, the proportion of cooperators remains $x_0$, and the average payoff also remains constant at the initial value. These experimental results validate that the initial proportion of cooperators does not affect the final evolutionary stable state, which is consistent with the theoretical analysis.

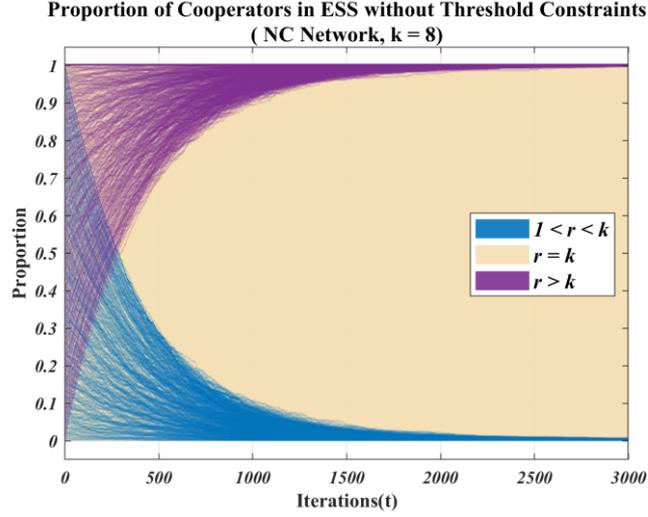

**Fig. 8** Proportion of cooperators over evolutionary iterations for different initial proportions of cooperators and three *r* values without threshold constraints. Blue line for $1 < r < k$, orange line for $r = k$, and purple line for $r > k$. The same applies to the following figure.

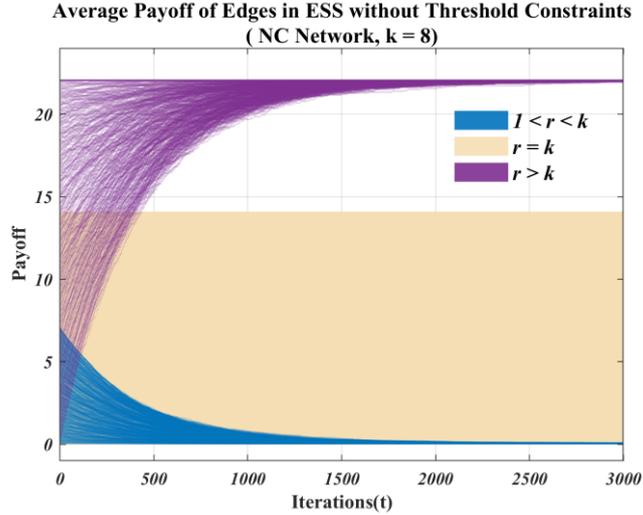

**Fig. 9** Average group payoff over evolutionary iterations for different initial proportions of cooperators and three r values without threshold constraints.

### 4.1.2 Cooperation Evolution Conditions Under Threshold Constraints

Under threshold constraints, the experimental setup is similar to the case without threshold constraints, with an additional parameter for the threshold $\theta$. By default, we set $\theta = k/2$, so here $\theta = 4$.

The first step is to investigate the effect of the parameter *r* on cooperation evolution. The range of *r* was set to $(1, 2.5k]$ with a step size of 0.1. Theoretical analysis suggests that the initial number of cooperators will influence the ESS, so we chose two extreme cases for $x_0$, namely $x_0 = 0$ and $x_0 = 1$, while keeping all other parameters the same as in **Fig. 7**. The results for ESS are shown in **Fig. 10**. In this figure, in addition to the values of the proportion of cooperators ($y_1$, blue line) and average population payoff ($y_2$, orange line) at ESS, we also include $y_1 = \theta/k$ (light blue dashed line), $r = k$ (black dashed line), and the piecewise fitting curve of $y_2$ when $r > k$ (purple and cyan dashed lines), as well as the ideal reference line for $y_2$ (green dashed line).

The results show that in edge-based games with threshold constraints, $r = k$ is also a critical point of evolutionary bifurcation. When $r \leq k$, both $y_1$ and $y_2$ are 0. However, after $r > k$, $y_1$ and $y_2$ are differ significantly from the case without threshold constraints: first, the changes in $y_1$ and $y_2$ are no longer linear, indicating some randomness in the evolution, and the stable state is no longer a fixed value. Secondly, for $x_0 = 0$, $y_1$ approaches but is less than 0.5, which is consistent with the theoretical conclusion that $y_1$ is close to but less than $\theta/k$. For $x_0 = 1$, two bifurcation points are observed: when $k < r < 2k$, the situation is similar to that for $x_0 = 0$, and when $r \geq 2k$, $y_1$ exceeds 0.5. Additionally, using **Eq.**(23), the ideal curve for $y_2$ is obtained ($y_2^{ideal} = r - 1$, green dashed line), and regardless of the value of $x_0$, the real value of $y_2$ is always lower than the ideal value, which aligns with the theoretical analysis. For $x_0=0$, the slope of the fitting curve for $y_2$ ($y_2 = 0.96r - 0.94$, purple dashed line) is less than that of the ideal curve, indicating that as $r$ increases, the gap between the real value and the ideal solution grows. For $x_0=1$, due to the two bifurcation points, the fitting curve for $y_2$ is a piecewise function: $y_2^1 = 0.97r - 1.05$ ($k < r < 2k$, purple dashed line) and $y_2^2 = 0.79r + 0.66$ ($r \geq 2k$, cyan dashed line), with slopes smaller than that of the ideal curve. Moreover, when $r \geq 2k$, the average payoff decreases significantly, confirming that for $x_0=1$, the larger the value of $r$, the greater the difference between the real value and the ideal solution.

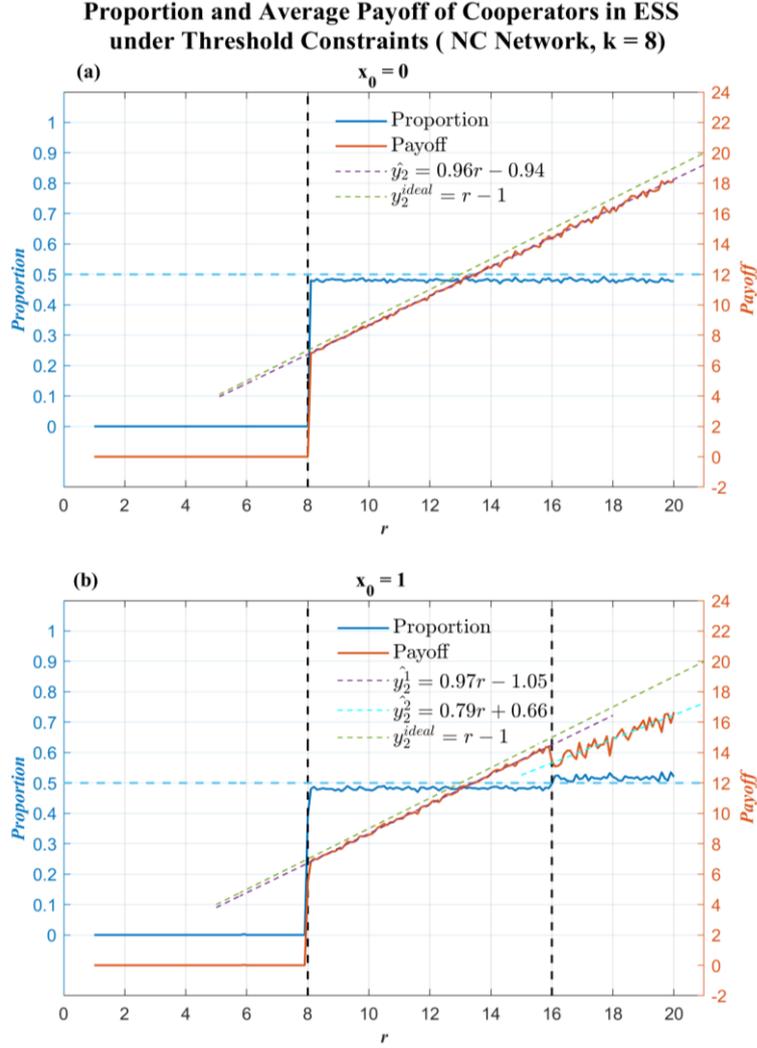

**Fig. 10** Proportion of cooperators (blue line) and average group payoff (orange line) in NC networks at ESS under threshold constraints for different $r$ values. Panels (**a**) and (**b**) correspond to initial cooperator proportions of 0 and 1, respectively. The black dashed lines represent reference lines for $r = k$ (or $2k$), and the light blue dashed line represents the reference line for $y_1 = \theta/k$. The purple (cyan) dashed lines show the fitted curves for payoff values when $r > 8$ (or $r > 16$), and the green dashed line represents the ideal payoff reference curve when $r > 8$.

In the second step, we compare the impact of the initial cooperator proportion (number) on the evolutionary stable state. Due to the threshold constraints, the ESS has a certain degree of randomness. Therefore, we perform 100 repeated experiments for 401 sets of initial cooperator number, recording the average, maximum, and minimum values of the cooperator proportion (**Fig. 11**, $y_1$) and the group average payoff (**Fig. 12**, $y_2$) at ESS. Based on the analysis in **Fig. 3**, simulations for different $r$-values reveal that the simulation results for three specific cases ($1 < r < k/\theta$, $r = k/\theta$, $k/\theta < r < k$) are identical (all yield $y_1 = 0$, $y_2 = 0$), so these three cases are combined as $1 < r < k$. For convenience, we define $r = (1+k)/2 = 4.5$, $r = k = 8$, $r = 1.5k = 12$, $r = 2k$

= 16, and $r = 2.5k = 20$ to represent five scenarios: $1 < r < k$ (orange), $r = k$ (purple), $k < r < 2k$ (green), $r = 2k$ (red), and $r > 2k$ (blue). In the figures, the x-axis represents the initial number of cooperators ($M \cdot x_0$), while the y-axis represents $y_1$ and $y_2$ for the ESS. The solid lines show the average $y_1$ and $y_2$ from the 100 repeated experiments, while the light-colored shaded area represents the range between the maximum and minimum values from those experiments. Additionally, in **Fig. 11**, the black line represents the ideal solution $y1 = \theta/k$, and the inset image is an enlarged display of the y-axis in the black dashed line region of the main image. In **Fig. 12**, the dashed line corresponds to the ideal solution for $y_2$ ($y_2^{ideal} = (2\theta r - k) \cdot cost / k = r - 1$).

From the analysis of the ESS cooperator proportion ($y_1$) in **Fig. 11**, we observe the following:

(1) $1 < r < k$: $y_1$ is unaffected by $M \cdot x_0$ and remains at 0. This means that the group will never cooperate under these conditions.

(2) $r = k$: $y_1$ is significantly influenced by $M \cdot x_0$. As $x_0$ increases, $y_1$ starts at 0 and increases almost linearly at first. However, the growth rate gradually decreases to 0, and $y_1$ stabilizes around 0.4 when $M \cdot x_0 > 250$.

(3) $k < r < 2k$: The impact of $M \cdot x_0$ on $y_1$ is smaller. The average $y_1$ stays around 0.48, and the maximum value of $y_1$ is always below the ideal value of $\theta/k$, maintaining "moderate cooperation".

(4) $r = 2k$: When $M \cdot x_0$ is around 140, the maximum value of $y_1$ exceeds the ideal solution of $\theta/k$. As $M \cdot x_0$ increases, $y_1$ continues to rise and surpasses the ideal value around $M \cdot x_0 = 225$. After reaching a steady state at $M \cdot x_0 = 285$, $y_1$ stabilizes between 0.48 and 0.51.

(5) $r > 2k$: The trend in $y_1$ is similar to the case when $r = 2k$. However, the critical points where $y_1$ surpasses the ideal solution and where the curve begins to stabilize occur at lower $M \cdot x_0$ values compared to $r = 2k$, and the stable $y_1$ values are significantly higher than those for $r = 2k$.

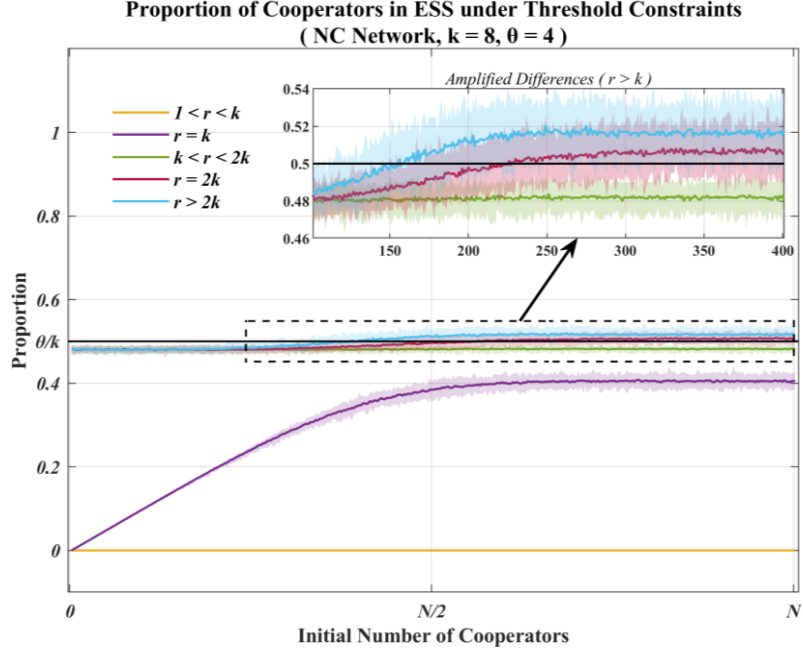

**Fig. 11** Under the threshold constraint, the cooperator proportion in the ESS across five different values of $r$ changes with the initial number of cooperators. The small inset in the figure displays a zoomed-in view of the vertical axis of the large plot's black dashed region. The black solid line represents the ideal cooperator proportion. Different colors represent the experimental results under various $r$ values: orange—$1 < r < k$, purple—$r = k$, green—$k < r < 2k$, red—$r = 2k$, blue: $r > 2k$. The solid lines represent the average cooperator proportion from 100 repeated experiments, and the light-colored shaded regions indicate the upper and lower limits corresponding to the maximum and minimum values from multiple experiments. This is the same as shown in **Fig. 12**.

The analysis of the average payoff of the ESS ($y_2$) in **Fig. 12**, as it varies with the initial number of cooperators, further supports the conclusions drawn from **Fig. 11**:

(1) $1 < r < k$: $y_2$ remains constantly zero regardless of $M \cdot x_0$, indicating that the group receives no benefit.

(2) $r = k$: $y_2$ is significantly influenced by $M \cdot x_0$. The payoff increases with $x_0$ but at a decelerating rate. After saturation, $y_2$ is lower than the ideal solution, and the difference between $y_2$ and the ideal solution is notably greater than the difference observed for the ESS when $k < r < 2k$.

(3) $k < r < 2k$: The variation in $y_2$ is minimal, with low data dispersion. The value of $y_2$ remains stable around 10.5, which is close to the ideal solution.

(4) $r = 2k$: Both the red and blue solid lines show a clear downward trend as $x_0$ increases. This indicates that as $x_0$ rises, "excessive cooperation" occurs in the edge games, leading to a decline in group payoff. Moreover, the larger the

$r$ value, the smaller the critical value of $x_0$ that causes "excessive cooperation," and the greater the difference between $y_2$ after "excessive cooperation" and the ideal solution.

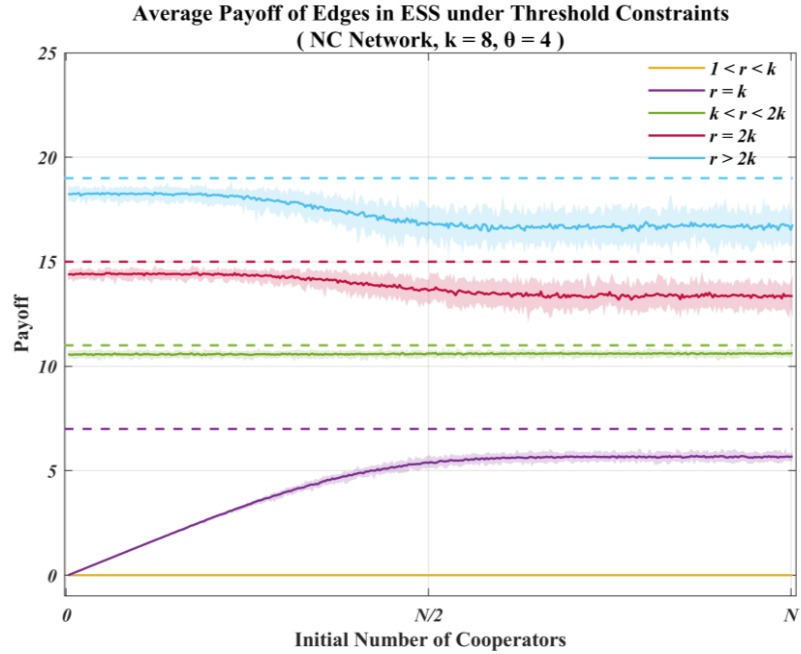

**Fig. 12** Average group payoff at ESS with varying initial cooperator numbers under threshold constraints. The dashed lines represent the theoretical optimal payoffs for each condition.

In summary, the simulation results align well with theoretical analysis: when $1 < r < k$, cooperative behavior cannot evolve under any condition; when $r = k$, both the cooperator proportion and group payoff at ESS show a near-logistic growth relationship with $x_0$, with an approximate deceleration phase in growth; when $k < r < 2k$, the ESS can consistently achieve "moderate cooperation" regardless of the value of $x_0$; when $r \geq 2k$, ESS can initially yield "moderate cooperation" when $x_0$ is small, however, beyond a critical threshold of $x_0$, the system tends to evolve into a state of "excessive cooperation", reducing group payoff, and the critical threshold $x_0$ decreases as $r$ increases.

Furthermore, the simulations confirm that for all $r$ values within the range $(k, 2k)$, except $r = 1.5k$, the system consistently satisfies $y_1 < \theta/k$, reinforcing that this condition is sufficient for achieving "moderate cooperation" in NC networks (figure omitted).

**4.2 Impact of Node Degree Distribution**

The discussions above are based on NC networks. However, when there is variability in node degrees within the network, the applicability of these conclusions requires further examination. Networks with heterogeneous node degrees can be divided into two categories: homogeneous networks and heterogeneous networks. In homogeneous networks, the degrees of all nodes are close to the average degree, such as in random networks and small-world networks. In heterogeneous networks, node degrees vary significantly, with some nodes having relatively small degrees and others having extremely large degrees. Networks with degree distributions that follow a power-law distribution are referred to as scale-free networks [34].

**4.2.1 Measures for Degree Heterogeneity**

There are several ways to define and measure the heterogeneity of a network. Here, we use two classic indicators for degree heterogeneity: the Gini coefficient [35] and the degree heterogeneity index H [36] to describe the heterogeneity of degree distribution in a network. They are defined as follows:

$$Gini = 1 - \sum_{i=1}^{N}\left[\frac{1}{N}\left(2\sum_{k=1}^{i}W_k - W_i\right)\right] \quad (24)$$

$$H = \langle k^2 \rangle / \langle k \rangle^2 \quad (25)$$

where:

- The Gini index is derived from the Lorenz curve in economics, and $W_i = d_i / \sum_{k=1}^{N} d_k$ measures the ratio of the degree value of the *i*-th node to the sum of the degree values of all nodes. The index ranges from 0 to 1, with higher values indicating greater heterogeneity. When the Gini index exceeds 0.5, the network is considered a scale-free network. Typically, networks with Gini values between 0.4 and 0.7 are classified as heterogeneous.
- The H index quantifies the dispersion of the degree distribution through the ratio of the second-order moment to the square of the first-order moment, where $\langle k \rangle$ is the average degree and $\langle k^2 \rangle$ is the second-order moment of the degree distribution. $H \geq 1$, and the higher the degree heterogeneity, the larger the H value.

## 4.2.2 Conditions for Cooperative Evolution in Complex Networks Without Threshold Constraints

In the case without threshold constraints, when there is variation in node degrees across the network, for the focus edges $e_m$, **Eqs.** (13) and (14) can be adjusted as follows:

$$\tau_{D \to C}(e_m) = \left[ \Pi_{e_m}^C(n_p+1, n_q+1) - \Pi_{e_m}^D(n_p, n_q) \right]_+ \\ = \left[ \frac{k_p(r-k_q) + k_q(r-k_p)}{k_p \cdot k_q} \cdot cost \right]_+, \forall e_m \in E, n_p < k_p, n_q < k_q \quad (26)$$

$$\tau_{C \to D}(e_m) = \left[ \Pi_{e_m}^D(n_p-1, n_q-1) - \Pi_{e_m}^C(n_p, n_q) \right]_+ \\ = \left[ \frac{k_p(k_q-r) + k_q(k_p-r)}{k_p \cdot k_q} \cdot cost \right]_+, \forall e_m \in E, n_p > 0, n_q > 0 \quad (27)$$

here, $k_{min}$ and $k_{max}$ represent the minimum and maximum degrees of the network, respectively.

For $\forall k_p, k_q$, $k_p \geq k_{min}$, $k_q \geq k_{min}$, $k_p \leq k_{max}$, and $k_q \leq k_{max}$ are true. Although it is not possible to obtain the macroscopic state by averaging the microscopic states as shown in **Eq.** (17), from the perspective of each edge:

(1) When $r > k_{max}$, **Eq.** (26) is positive, while **Eq.** (27) equals zero, $\tau_{D \to C}(e_m) > \tau_{C \to D}(e_m)$, indicating that all edges in a state of defection will eventually evolve into cooperation. The ESS is C.

(2) When $r < k_{min}$, **Eq.** (26) is zero, while **Eq.** (27) is positive, $\tau_{D \to C}(e_m) < \tau_{C \to D}(e_m)$, indicating that all cooperative edges will evolve into defection. The ESS is D.

When the ESS is C, we have $n_p = k_p$, $n_q = k_q$. From **Eq.** (3), we obtain:

$$\overline{\Pi}_{e_m} = 2(r-1) \cdot cost \quad (28)$$

When the ESS is D, we have $n_p = n_q = 0$. From **Eq.** (6), we obtain: $\overline{\Pi}_{e_m} = 0$.

It is evident that the average group payoff is the same as that in NC networks. Based on the discussion above, it can be concluded that: in the absence of threshold constraints, the sufficient condition for group cooperation in complex networks with non-uniform degree distributions is $r > k_{max}$, and the sufficient condition for group defection is $r < k_{min}$.

Simulations were performed with ER random networks, WS small-world networks, and BA scale-free networks, all having the same number of nodes (100) and edges (400) as NC networks. The initial conditions were set as $x_0 = 0$ and $x_0 = 1$, with the *cost* set to 1. The value of $r$ was chosen from 196 equally spaced values in the interval [1, 40] for evolutionary simulations. The resulting cooperative ratio and average group payoff as functions of $r$ are shown in **Fig. 13**. Each column corresponds to a different network type, and each row corresponds to a different initial cooperative ratio. In the subfigures, the solid lines have the same meaning as in **Fig. 7**, and the black dashed lines represent $r = k_{min}$ and $r = k_{max}$.

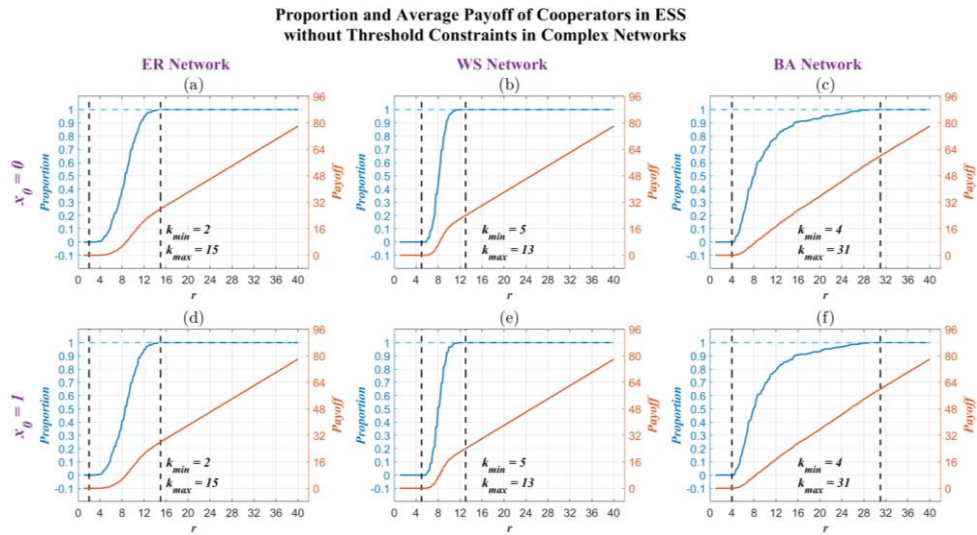

**Fig. 13** Proportion of cooperators (blue lines) and average group payoff (orange lines) at evolutionary equilibrium for ER networks (**a**, **d**), WS networks (**b**, **e**), and BA networks (**c**, **f**) as a function of $r$. Panels (**a–c**) and (**d–f**) correspond to initial proportions of cooperators of 0 and 1, respectively. Black dashed lines represent the reference lines $r = k_{min}$ and $r = k_{max}$, and light blue dashed lines represent the reference line $y_1 = 1$.

The results show that, in the absence of threshold constraints, all three network types satisfy the following conditions under both initial conditions: when $r > k_{max}$, $y_1$ is 1 and $y_2$ is $2r-2$; when $r < k_{min}$, $y_1$ is 0 and $y_2$ is 0; for the values of $r$ where $y_1 = 0$ or $y_1 = 1$, there is no difference between $x_0 = 0$ and $x_0 = 1$; for the values of $r$ where $0 < y_1 < 1$, there is a slight difference between $x_0 = 0$ and $x_0 = 1$. Therefore, when exploring the conditions for group cooperation, the initial conditions are not the key factor. It is important to note that in **Fig. 13**(a) and **Fig. 13**(d), when $k_{min} \leq r \leq 3.4$, $y_1 = 0$ still holds, and in **Fig. 13**(c) and **Fig. 13**(f), when $29 \leq r \leq k_{max}$, $y_1 = 1$ still holds. Hence, the conditions $r < k_{min}$ and $r > k_{max}$ are not necessary and sufficient conditions; they are merely sufficient conditions.

## 4.2.3 Evolutionarily Stable Strategies of Homogeneous Networks under Threshold Constraints

Section 4.2.2 provides a preliminary discussion on edge games without threshold constraints. The conclusions derived when there are differences in node degrees can be appropriately inferred from the theoretical solutions of NC networks. When threshold constraints are present, non-NC networks, due to the increased degree heterogeneity, exhibit stronger randomness in the asynchronous update process. As a result, theoretical reasoning becomes more challenging. Therefore, the next two subsections aim to alter the network heterogeneity and conduct an exploratory analysis of the conditions for "moderate cooperation" in networks.

In this section, we refer to the NC network experiments in **Fig. 10**, maintaining the number of nodes as 100 and the number of edges as 400. The degree sequence is gradually expanded by increasing the degree interval, such as [8], [7, 8, 9], [6, 7, 8, 9, 10], and so on, while ensuring that the probability of each node having a specific degree remains as uniform as possible (excess nodes are assigned a degree value of the median, which is 8). Using the configuration model [37], self-loops and duplicate edges are adjusted, and five homogeneous networks with different degree distributions are generated, as shown in **Fig. 14(I)–(V)**. From the Gini index and H-index, it is evident that the heterogeneity of these five networks gradually increases, but the Gini index remains below 0.2, indicating low heterogeneity.

For the five homogeneous networks mentioned above, network evolution is conducted with the same variable settings as in **Fig. 13**. The ESS of the networks, including $y_1$ and $y_2$ as functions of $r$, are shown in **Fig. 14(a)–(j)**. Compared to **Fig. 10** and **Fig. 13**, **Fig. 14** introduces a third axis (green), which records the number of nodes exceeding the threshold constraint in the evolved stable cooperation network (denoted as $y_3$), to assess whether the network's stable state exhibits "excessive cooperation". Section 4.1.2 precisely defines the conditions for "moderate cooperation" as $k < r < 2k$ in NC networks, while Section 4.2.2 fixes the critical values for full defection and full cooperation as $r = k_{min}$ and $r = k_{max}$, respectively. Based on the previous discussions, **Fig. 14** adds reference lines for $r = k_{min}$ and $r = k_{max}$ (black dashed lines), as well as for $r = 2k_{min}$ and $r = 2k_{max}$ (yellow dashed lines), to explore the boundary conditions for "moderate cooperation".

From **Fig. 14(a)**–(**e**), it can be seen that when $x_0 = 0$, the critical $r$-values for the jump in $y_1$ are similar to those in **Fig. 13**. When $r < k_{min}$, $y_1 = 0$, and when $r > k_{max}$, , group cooperation becomes stable, with $y_1$ approaching the ideal value $y_1 = \theta/k$, $y_2$ approaching 0, and $y_3$ remaining at 0. The evolutionary stable state is "moderate cooperation." When $x_0 = 1$, as $r$ increases, all networks exhibit "excessive cooperation" ($y_3 > 0$). In **Fig. 14(f)**–(**j**), the green text highlights the continuous intervals where $y_3=0$. It is clear that when $2k_{min} > k_{max}$ (*Network* 1-3, **Fig. 14(f)**–(**h**)), the "moderate cooperation" interval lies precisely between the second black dashed line and the first yellow dashed line, that is, $k_{max} < r < 2k_{min}$. In this case, $y_1$ no longer increases and $y_3=0$. When $2k_{min} \leq k_{max}$ (**Fig. 14(i)**–(**j**)), the "moderate cooperation" interval becomes unstable. For example, in *Network* 4, the "moderate cooperation" interval is $k_{max} < r < 14$, and in *Network* 5, there is no "moderate cooperation" interval. As network heterogeneity increases, the condition $2k_{min} \leq k_{max}$ becomes more common, and the initial proportion of cooperators introduces uncertainty into the cooperative evolution outcomes, which is detrimental to achieving a stable cooperative structure (for detailed analysis, see Appendix C, **Fig.S1**). Subsequent experiments also show that when the network heterogeneity is sufficiently large, even if the initial proportion of cooperators is 0, it is still impossible to obtain an $r$-value for "moderate cooperation" (**Fig.S2**).

The above exploratory analysis can only yield one definitive conclusion: when network heterogeneity is low and $2k_{min} > k_{max}$, the sufficient condition for "moderate cooperation" in the network is $k_{max} < r < 2k_{min}$. When network heterogeneity is high, the conditions for the evolution of "moderate cooperation" require further discussion.

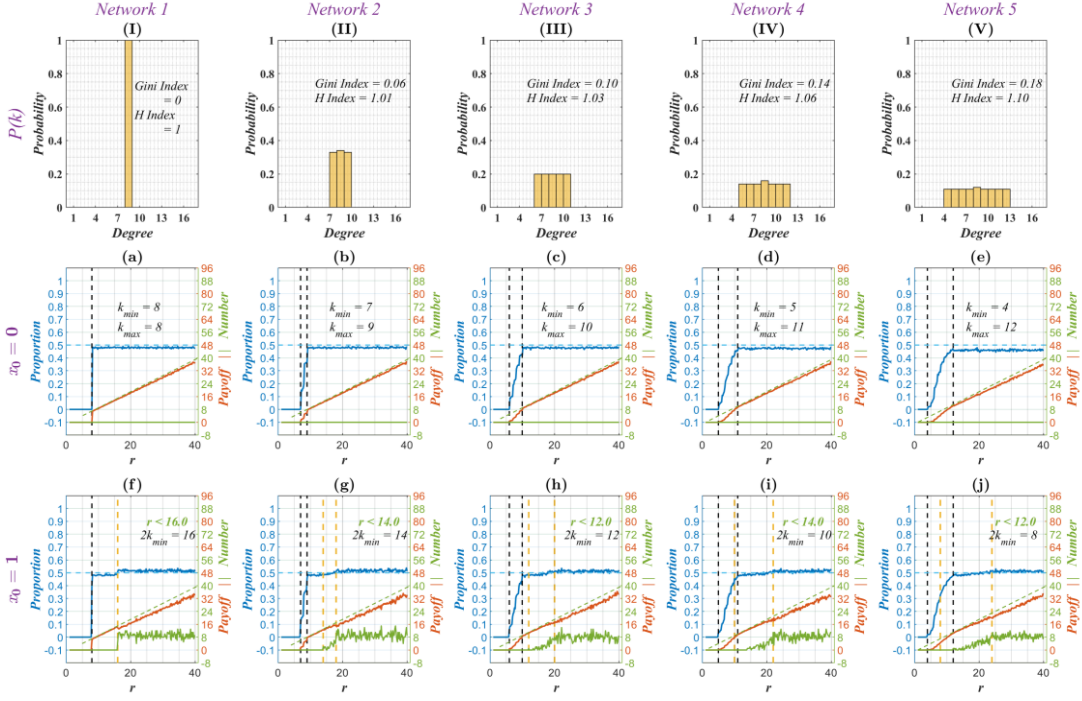

**Fig. 14** Degree distribution of five homogeneous networks (**I** to **V**) with the same number of nodes and edges, and the changes in the proportion of cooperators (blue lines), average group payoff (orange lines), and the number of nodes exceeding the threshold constraints (green lines) at ESS as a function of $r$ (**a** to **j**) under threshold constraints. The degree heterogeneity of the five networks increases from columns 1 to 5. The initial proportions of cooperators in **a–e** and **f–j** are 0 and 1, respectively, and the range of $r$ values for $y_3 = 0$ is marked in **f–j**. Black dashed lines represent reference lines $r = k_{min}$ and $r = k_{max}$, yellow dashed lines represent reference lines $r = 2k_{min}$ and $r = 2k_{max}$, light blue dashed lines represent the reference line $y_1 = \theta/k$, and green dashed lines represent $y_2^{ideal} = r - 1$.

### 4.2.4 Evolutionarily Stable Strategies of Heterogeneous Networks under Threshold Constraints

To understand the impact of degree heterogeneity on the evolution of cooperation in edge-based games, it is essential to generate networks with controllable and measurable degree heterogeneity for conducting comparative experiments. Based on the modified Weibull distribution degree distribution function (**Eq.** (29)) proposed by Ozbay and Nguyen [38] and combined with the configuration model [37], we generated heterogeneous networks with a continuous degree heterogeneity parameter ($\sigma$), where the number of nodes (N = 100) and edges (M = 400) are fixed. Since the network size is relatively small, self-loops and duplicate edges inevitably occur during the network configuration process. To address this, such edges are removed and reconnected to other nodes with similar degree values to the original nodes. This ensures that the degree distribution of the generated network closely matches the given Weibull distribution. The final degree distribution of the network is shown in **Fig. 15**.

$$f(x;k,\sigma) = \frac{-\ln(\sigma)}{\langle k \rangle}(\frac{x}{\langle k \rangle})^{-\ln(\sigma)-1}\exp(-(x/\langle k \rangle)^{-\ln(\sigma)}); x \geq 0; \sigma \in (0,1); \langle k \rangle \in \mathbb{R}^+ \quad (29)$$

here, the modified Weibull distribution has only two parameters: the average degree $\langle k \rangle$ (set to 8 for all experiments in this section) and $\sigma$, which controls the degree distribution's heterogeneity. Several typical values of $\sigma$ were chosen: $\sigma = 0$ for a constant distribution, $\sigma = 0.03$ for a Gaussian distribution, $\sigma = 0.14$ for a Rayleigh distribution, $\sigma = 0.37$ for an exponential distribution, and $\sigma > 0.37$ for a long-tail distribution (corresponding to a scale-free network). Since the networks used in the experiments are not large-scale, we selected $\sigma$ values of 0.4, 0.45, 0.5, and 0.55 to model the degree distribution of scale-free networks in the comparative experiments. It is evident that as $\sigma$ increases, the network's heterogeneity significantly increases, whether measured by the Gini index or the H-index.

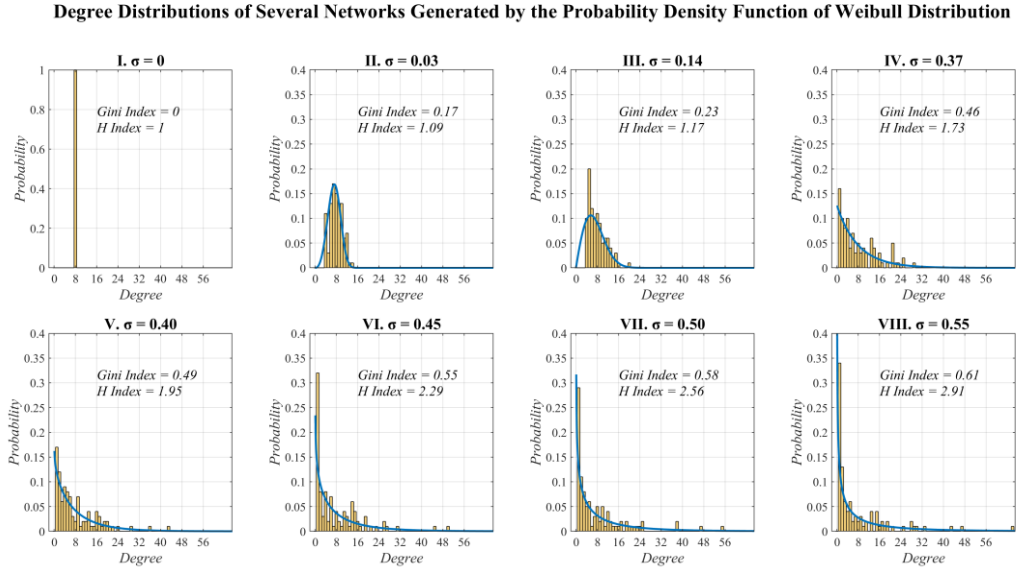

**Fig. 15** Illustration of degree distributions for multiple types of networks derived from the Weibull probability density distribution (the blue solid line represents the function curve generated based on the $\sigma$ parameter, and the yellow striped area represents the actual degree distribution of the generated network).

As discussed in Section 4.2.3, highly heterogeneous networks are prone to the condition $2k_{min} \leq k_{max}$. In **Fig. 15**, all networks except (**I**) exhibit this characteristic. In such networks, the value of *r* for "moderate cooperation" evolution is highly influenced by the initial proportion of cooperators, making it difficult to determine a specific range of *r* values for achieving cooperative evolution. Experimental results shown in **Fig.S2** further demonstrate that when $x_0 = 0$ and $\sigma \geq 0.37$, it becomes impossible to find a definitive *r* value that both maintains $y_1$ at a stable high value and controls $y_3=0$. Therefore, highly heterogeneous networks cannot adapt the value of *r* through the reduction of $x_0$ to promote cooperative evolution.

In real-world scenarios, many networks are highly heterogeneous, where the degree values of the nodes connected by edges often differ significantly, resulting in large variations in the size of the game groups involved in edge games. When nodes have cooperation threshold constraints, a fixed *r* value cannot simultaneously satisfy the "moderate cooperation" requirement for both sides of the game group.

To address this issue, we propose a variable synergy factor (*r*) edge-based game model that adapts to networks with various degree distributions. This approach is simple and feasible. As shown in **Fig. 2**, each edge participates in a game group centered around the nodes on both sides, with a determined group size (denoted as $k_p$ and $k_q$). Instead of using a single *r*-value for the entire network, we assign each game group a specific *r*-value. If the game groups involved in each focal edge can achieve "moderate cooperation", then the entire network can naturally achieve "moderate cooperation". Here, we introduce a new variable, *n-fold*, as the multiple of the *r*-value relative to the node degree, adjusting the *r*-value to change in proportion to the degree *k*. For a game group centered on node $v_x$ (with degree $k_x$), we define:

$$r_x = \textit{n-fold} \cdot k_x, \quad x = 1, \ldots, N \tag{30}$$

Then, the payoff for $e_m$ is given by:

$$\Pi^C_{e_m}(n_p, n_q) = \frac{n_p \cdot r_p \cdot cost}{k_p} + \frac{n_q \cdot r_q \cdot cost}{k_q} - 2 \cdot cost = \left[\textit{n-fold} \cdot (n_p + n_q) - 2\right] \cdot cost \tag{31}$$

$$\Pi^D_{e_m}(n_p, n_q) = \frac{n_p \cdot r_p \cdot cost}{k_p} + \frac{n_q \cdot r_q \cdot cost}{k_q} = \textit{n-fold} \cdot (n_p + n_q) \cdot cost \tag{32}$$

The improved payoff function no longer includes the node degree value, and is only influenced by the number of cooperators on both sides. Therefore, the impact of heterogeneity in the network is reduced.

Based on this new payoff function, we explore the impact of the *n-fold* parameter on cooperation evolution. The value of *n-fold* is set to range from [0.9, 2.1] with a step size of 0.1, covering various possibilities from $r_x < k_x$ to $r_x > 2k_x$. Similar to **Fig. 14**, we compare the ESS values of $y_1$, $y_2$, and $y_3$ for eight networks at different *n-fold* values (**Fig. 16**). It can be observed that in all subplots, for *n-fold* within the range of [1.1, 1.9], $y_1$ no longer increases but remains relatively stable, and maintains $y_3 = 0$, corresponding to $k_x < r_x < 2k_x$.

Since $y_1$ is relatively stable in the range of [1.1, 1.9], we calculate the mean value of $y_1$ ($\bar{y}_1$) within this range. For homogeneous networks (**Fig. 15(I)–(III)**), $\bar{y}_1$ still approaches the reference line $\theta/k$. As the network heterogeneity increases, the value of $\bar{y}_1$ gradually decreases, indicating that for networks with the same number of nodes and edges, as network heterogeneity increases, the number of cooperative edges in the cooperation structure will significantly decrease. Drawing an analogy to UAV swarm networks, it can be inferred that for a network with fixed communication resources, the more uniform the edge distribution in the network's underlying topology, the higher the network's collaboration efficiency.

A vertical comparison of the two extreme cases, $x_0 = 0$ and $x_0 = 1$, shows the following: When $x_0 = 0$, there is only one jump at *n-fold* = 1, and for *n-fold* > 1, $y_1$ tends to stabilize. In contrast, when $x_0 = 1$, there is an additional jump at *n-fold* = 2, and for *n-fold* ≥ 2, $y_1$ continues to increase, $y_3 > 0$, leading to a situation of "excessive cooperation". Moreover, for all networks except the regular network (**Fig. 15(I)**), the value of $\bar{y}_1$ at $x_0 = 1$ is slightly larger than that at $x_0 = 0$. Given the inherent randomness in the experiments, we selected some specific *n-fold* values and performed 100 repeated trials for both $x_0 = 0$ and $x_0 = 1$ (see Appendix C, **Fig.S3**). The results show that the randomness in the experiments can account for the difference in values between the two cases. The difference is not significant, and for different networks, the value of $\bar{y}_1$ for $x_0 = 1$ are sometimes higher and sometimes lower than those for $x_0 = 0$.

From the discussion above, it is clear that the variable synergy factor edge-based game model enables networks with any degree distribution to achieve "moderate cooperation." For any initial proportion of cooperators, the necessary and sufficient condition for "moderate cooperation" is $r_x = n\text{-}fold \cdot k_x$, where $n\text{-}fold \in (1, 2)$.

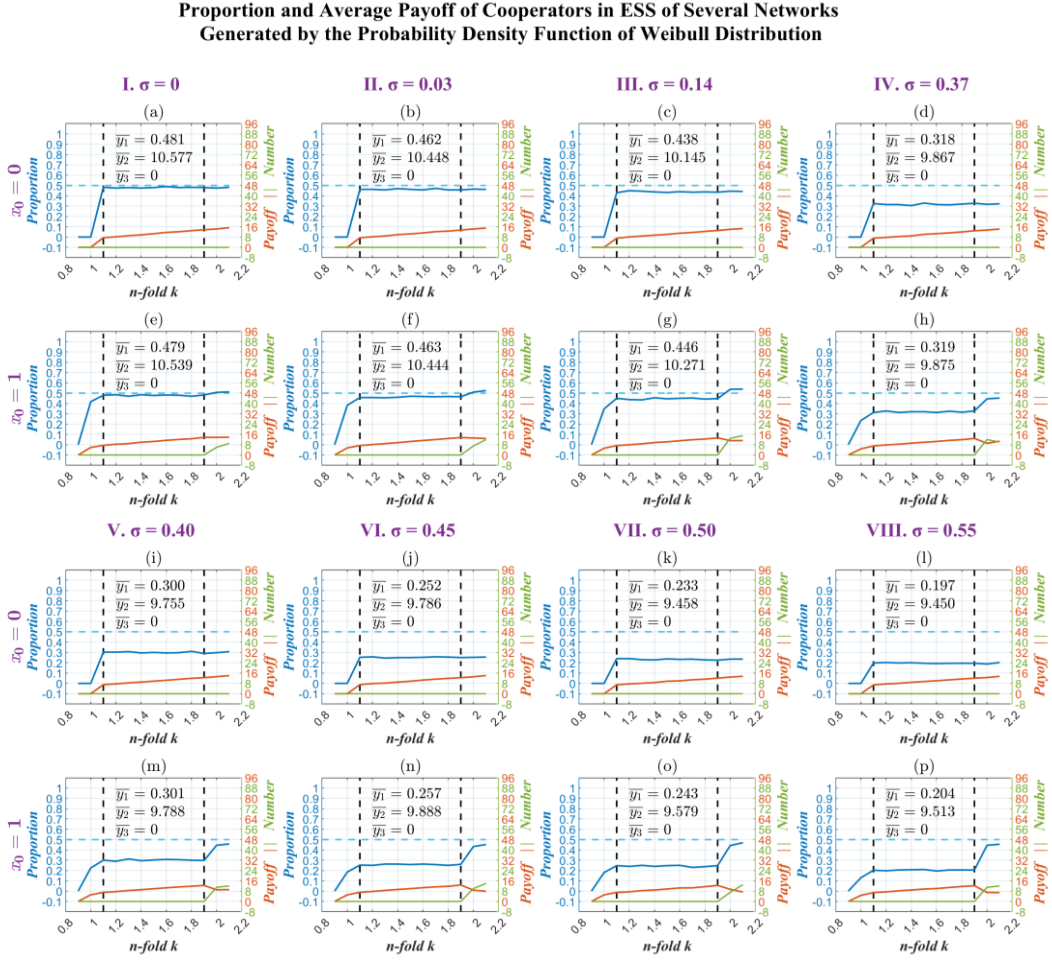

**Fig. 16** Proportion of cooperators (blue lines) and average group payoff (orange lines) at ESS for eight networks with the same number of nodes and edges, based on the Weibull probability density distribution, as a function of $n$-fold under threshold constraints. Networks **I** to **VIII** correspond to increasing degree heterogeneity. Panels (**a–d**) and (**i–l**) have an initial proportion of cooperators of 0, while panels (**e–h**) and (**m–p**) have an initial proportion of cooperators of 1. Black dashed lines represent reference lines for $n\text{-}fold = 1.1$ and $n\text{-}fold = 1.9$, and the three values in each subplot represent the average values of $y_1$, $y_2$, and $y_3$ within the interval $n\text{-}fold = [1.1, 1.9]$. Light blue dashed lines represent the reference line $y_1 = \theta/k$.

### 4.3 Comparison with Other Cooperation Algorithms

The selection of cooperative partners under threshold constraints proposed in our study is similar to the degree-restricted matching problem in traditional network optimization. Unlike traditional optimization algorithms, we utilize edge games for network cooperative partner selection, which is based on the realistic situation of agents with bounded rationality and limited information. Agents spontaneously choose cooperative partners through interactions with their neighbors, seeking the optimal cooperative structure of the network. This process fully reflects the intelligence of individuals. Here, we select the classic Linear Programming (LP) algorithm as the baseline, aiming to maximize the number of cooperative edges. We also choose the Greedy (GR) algorithm, Local Search (LS) algorithm, Particle Swarm Optimization (PSO), and Genetic Algorithm (GA) as comparison algorithms. We contrast these algorithms with the Edge Game (EG) model with a variable synergy factor (set *n-fold* to 1.1) to explore the advantages and disadvantages of the edge game algorithm.

The networks used in this experiment are consistent with those in Section 4.2.3. For networks of this scale, the LP method can directly find the exact solution; however, the algorithm's complexity is too high (exponential). Therefore, we use the cooperative edge proportion obtained from the LP method as a benchmark and calculate the ratio of other algorithms to the benchmark. The standardized ratio values are visualized in **Fig. 17**. The line in **Fig. 17** represents the cooperative edge proportion obtained by the LP algorithm. It is evident that for networks with low heterogeneity (**Fig. 15(I)-(III)**), the ideal cooperative proportion is 0.5 ($\theta/k$), but as heterogeneity increases, the ideal value of $y_1$ decreases.

Comparing the performance of various algorithms, we find that the worst-performing algorithm is GR. LS, PSO, and GA all use GR's solution as the initial solution for optimization. Despite this, the experimental results show that the EG algorithm (red bars) performs similarly to several optimization algorithms, with a slight advantage. Moreover, the EG model with a varying *r*-value is minimally influenced by the initial proportion of cooperators, and its optimization performance is stable.

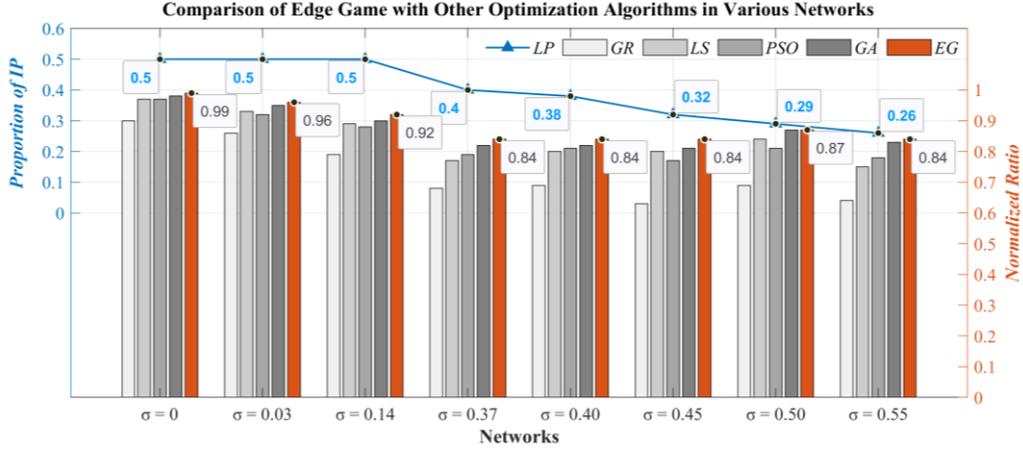

**Fig. 17** Comparison of the proportions of cooperators and ideal solution ratios obtained by four optimization algorithms and the EG algorithm in different networks. The x-axis represents networks with gradually increasing heterogeneity. The left y-axis shows the ideal solution obtained by the LP algorithm, represented by the blue line in the figure. The right y-axis shows the normalized ratio of the comparison algorithms to the ideal solution. The red bars represent the EG algorithm

In addition to the slightly better performance of the EG, the biggest advantage of EG is its relatively low algorithmic complexity. As shown in **Table 2**, the time complexity of EG (O($t \cdot N \cdot M$)) is only slightly higher than that of GR, slightly lower than that of LR, PSO, and GA, and significantly lower than that of IP.

**Table 2** Time Complexity of Algorithms (Big-O Notation, $t$ is the number of iterations)

| Algorithm | IP | GR | LS | PSO | GA | EG |
|---|---|---|---|---|---|---|
| Complexity | O($2^M$) | O(M) | O($t \cdot M^2$) | O($t \cdot Size \cdot M^2$), where Size is the particle swarm size | O($t \cdot Size \cdot M^2$), where Size is the population size | O($t \cdot N \cdot M$) |

Considering both algorithm performance and time complexity, EG proposed in our study has significant advantages.

## 5. Conclusion and Outlook

Our study innovatively proposes an evolutionary game model that can determine cooperative partners—Edge Game. This model is simple and universal, using the edges on complex networks as virtual game objects. Based on the Smith dynamics process and within the framework of group games, by configuring the synergy factors to meet the network's "moderate cooperation" conditions, a stable cooperative structure can be achieved for any network at the evolutionary stable state.

Following the traditional research approach of network evolutionary games, this study gradually increases the constraints and network heterogeneity, exploring the

conditions for cooperation evolution in edge games. The main conclusions are as follows:

(1) Combining macroscopic mean-field theory and numerical simulations, we theoretically derive and experimentally verify the conditions for network cooperation evolution without threshold constraints on the number of cooperators: The necessary and sufficient condition for cooperation evolution in NC networks is $r > k$, and the sufficient condition for cooperation evolution in complex networks with non-uniform degree values is $r > k_{max}$. For any initial strategy distribution, as long as the synergy factor meets these conditions, all edges in the network can be transformed into cooperative edges.

(2) When there is a threshold ($\theta$) for the number of cooperations at each node, the network's evolution is influenced by the synergy factors and initialization strategies. Stable cooperation states exist in two types of equilibrium:

a) Moderate cooperation: When cooperation edges are as high as possible under the threshold constraint, individual rationality and collective rationality reach a win-win situation.

b) Excessive cooperation: When the synergy factor is too high, selfishness among the game participants causes the proportion of cooperation edges to become abnormally high. Some nodes become overloaded with cooperation, losing their ability to cooperate. The average group payoff decreases, and collective rationality cannot be maintained.

Therefore, for different types of networks, we determine the conditions for "moderate cooperation," as follows:

a) NC networks: For $\forall x_0 \in [0,1]$, the "moderate cooperation" condition is $k < r < 2k$, with the proportion of cooperators is slightly less than the ideal value $\theta/k$, and the upper bound of the group average payoff is $(2\theta r - k) \cdot cost / k$.

b) Homogeneous networks satisfying $2k_{min} > k_{max}$: For $\forall x_0 \in [0,1]$, the sufficient condition for "moderate cooperation" is $k_{max} < r < 2k_{min}$.

c) Any network (applicable to regular, homogeneous, and heterogeneous networks): Using the edge game model with a variable $r$-value, for $\forall x_0 \in [0,1]$, the necessary and sufficient condition for "moderate cooperation" in any network is $r_x = \textit{n-fold} \cdot k_x$, $\textit{n-fold} \in (1, 2)$.

(3) As network heterogeneity increases, under threshold constraints, the proportion of cooperators at the ESS of edge games shows a downward trend. The network structure with the highest proportion of cooperators is the NC network.

A key feature of this algorithm is that it adopts an evolutionary game perspective and derives the evolutionary stable state from swarm intelligence. Compared to other optimization-based cooperation algorithms, the edge game algorithm strikes a balance between algorithm performance and time complexity. Additionally, its advantages, including limited rational decision-making, local information usage, and dynamic self-organization mechanisms, make it more adaptable to complex interaction scenarios than global optimization algorithms.

It is important to note that the edge game model presented in this paper represents its most basic and fundamental form, aimed at initially solving the cooperation partner selection and optimization configuration problem. In practical applications, the edge game model is highly adaptable and can be modified based on real-world requirements. For example, in heterogeneous information networks with node attribute differences, the payoff matrix parameters can be altered to achieve cooperation optimization between heterogeneous nodes. In edge-heterogeneous networks, additional constraints such as edge attributes can be added, utilizing edge weights, geographical distances, and prioritizing nodes, defining familiarity, or incorporating reputation mechanisms and penalty mechanisms to adjust the game model. For multi-agent cooperation problems, a hyper-edge-based edge game model can be designed to address complex tasks involving multi-agent collaboration. These are potential directions for future research breakthroughs.

**References**

[1] Ö. Bodin, M. Mancilla García, G. Robins, Reconciling conflict and cooperation in environmental governance: A social network perspective, Annu. Rev. Environ. Resour. 45 (2020) 471–495. https://doi.org/10.1146/annurev-environ-011020-064352.

[2] J. Kondo, D. Li, D. Papanikolaou, Trust, Collaboration, and Economic Growth,

Manage. Sci. 67 (2021) 1825–1850. https://doi.org/10.1287/mnsc.2019.3545.

[3] Y. Zheng, Evolutionary game analysis on the cross-organizational cooperative R&D strategy of general purpose technologies under two-way collaboration, Technol. Soc. (2024). https://doi.org/10.1016/j.techsoc.2024.102473.

[4] J. M. Smith, G. R. Price, The logic of animal conflict, Nature 246 (1973) 15–18.

[5] J.M. Pacheco, F.C. Santos, Co-evolution of risk and cooperation in climate policies under wealth inequality, PNAS Nexus 3 (2024) pgae550. https://doi.org/10.1093/pnasnexus/pgae550.

[6] M. Saeedian, C. Tu, F. Menegazzo, P. D'Odorico, S. Azaele, S. Suweis, Modelling co-evolution of resource feedback and social network dynamics in human-environmental systems, New J. Phys. 26 (2024) 83004. https://doi.org/10.1088/1367-2630/ad67fe.

[7] Y. Xu, G. Gui, H. Gacanin, F. Adachi, A Survey on Resource Allocation for 5G Heterogeneous Networks: Current Research, Future Trends, and Challenges, IEEE Commun. Surv. Tutorials 23 (2021) 668–695. https://doi.org/10.1109/COMST.2021.3059896.

[8] Garrett Hardin, The tragedy of the commons, Science (1968) 1243–1248. https://doi.org/10.1126/science.162.3859.1243.

[9] C. Tu, P. D'Odorico, Z. Li, S. Suweis, The emergence of cooperation from shared goals in the governance of common-pool resources, Nat Sustain 6 (2022) 139–147. https://doi.org/10.1038/s41893-022-01008-1.

[10] J. M. Smith, Evolution and the Theory of Games, Cambridge University Press, Cambridge, 1982.

[11] M.A. Nowak, Five rules for the evolution of cooperation, Science 314 (2006) 1560–1563.

[12] Wei H., Zhang J., Zhang C., Evolutionary dynamics of direct and indirect reciprocity on networked populations, Swarm Evol. Comput. 88 (2024) 101611. https://doi.org/10.1016/j.swevo.2024.101611.

[13] F.C. Santos, M.D. Santos, J.M. Pacheco, Social diversity promotes the emergence of cooperation in public goods games, Nature 454 (2008) 213–216. https://doi.org/10.1038/nature06940.

[14] M.A. Nowak, R.M. May, Evolutionary games and spatial chaos, Nature 359 (1992) 826–829. https://doi.org/10.1038/359826a0.

[15] C. Hauert, M. Doebeli, Spatial structure often inhibits the evolution of cooperation in the snowdrift game, Nature 428 (2004) 643–646. https://doi.org/10.1038/nature02360.

[16] H. Ohtsuki, C. Hauert, E. Lieberman, M.A. Nowak, A simple rule for the evolution of cooperation on graphs and social networks, Nature 441 (2006) 502–505. https://doi.org/10.1038/nature04605.

[17] B. Allen, G. Lippner, Y.-T. Chen, B. Fotouhi, N. Momeni, S.-T. Yau, M.A. Nowak, Evolutionary dynamics on any population structure, Nature 544 (2017) 227–230. https://doi.org/10.1038/nature21723.

[18] Y.P. Kuo, C. Nombela-Arrieta, O. Carja, A theory of evolutionary dynamics on any complex population structure reveals stem cell niche architecture as a spatial suppressor of selection, Nat. Commun. 15 (2024) 4666. https://doi.org/10.1038/s41467-024-48617-2.

[19] A. Li, L. Zhou, Q. Su, S.P. Cornelius, Y.-Y. Liu, L. Wang, S.A. Levin, Evolution of cooperation on temporal networks, Nat. Commun. 11 (2020) 2259. https://doi.org/10.1038/s41467-020-16088-w.

[20] A. Civilini, O. Sadekar, F. Battiston, J. Gómez-Gardeñes, V. Latora, Explosive cooperation in social dilemmas on higher-order networks, Phys. Rev. Lett. 132 (2024) 167401. https://doi.org/10.1103/PhysRevLett.132.167401.

[21] U. Alvarez-Rodriguez, F. Battiston, G.F. de Arruda, Y. Moreno, M. Perc, V. Latora, Evolutionary dynamics of higher-order interactions in social networks, Nature Human Behaviour 5 (2021) 586–595. https://doi.org/10.1038/s41562-020-01024-1.

[22] A. Sheng, Q. Su, L. Wang, J.B. Plotkin, Strategy evolution on higher-order networks, Nat. Comput. Sci. 4 (2024) 274–284. https://doi.org/10.1038/s43588-024-00621-8.

[23] J. Díaz, D. Mitsche, A survey of the modified Moran process and evolutionary graph theory, Computer Science Review 39 (2021) 100347. https://doi.org/10.1016/j.cosrev.2020.100347.

[24] C. Zhang, Q. Li, Y. Zhu, J. Zhang, Dynamics of task allocation based on game theory in multi-agent systems, IEEE Trans. Circuits Syst. II 66 (2019) 1068–1072. https://doi.org/10.1109/TCSII.2018.2873006.

[25] A. Yaman, J.Z. Leibo, G. Iacca, S. Wan Lee, The emergence of division of labour through decentralized social sanctioning, Proc. R. Soc. B: Biol. Sci. 290 (2023) 20231716. https://doi.org/10.1098/rspb.2023.1716.

[26] J.R. Marden, J.S. Shamma, Game theory and control, Annu. Rev. Control Rob. Auton. Syst. 1 (2018) 105–134. https://doi.org/10.1146/annurev-control-060117-105102.

[27] H. Wu, H. Deng, J. Li, H. Luo, Research on Task Collaboration Over Heterogeneous Networks Based on Evolutionary Game Theory, in: 2024 IEEE International Conference on Systems, Man, and Cybernetics (SMC), IEEE, Kuching, Malaysia, 2024: pp. 4472–4479. https://doi.org/10.1109/SMC54092.2024.10831853.

[28] J. Fan, H. Du, G. Li, X. He, The effect of multi-tasks mechanism on cooperation in evolutionary game, Chaos: Interdiscip. J. Nonlinear Sci. 34 (2024). https://doi.org/10.1063/5.0210787.

[29] C. Sun, X. Wang, H. Qiu, Q. Zhou, Game theoretic self-organization in multi-satellite distributed task allocation, Aerospace Science and Technology 112 (2021) 106650. https://doi.org/10.1016/j.ast.2021.106650.

[30] H. Whitney, Congruent graphs and the connectivity of graphs, Am. J. Math. 54 (1932) 150–168. https://doi.org/10.2307/2371086.

[31] T.S. Evans, R. Lambiotte, Line graphs, link partitions, and overlapping communities, Phys. Rev. E 80 (2009) 16105. https://doi.org/10.1103/PhysRevE.80.016105.

[32] W.H. Sandholm, Revision Protocols and Evolutionary Dynamics, in: Population Games and Evolutionary Dynamics, The MIT Press, London, England, 2010: pp. 119–138.

[33] Michael J. Smith, The stability of a dynamic model of traffic assignment---an application of a method of Lyapunov, Transportation Science 18 (1984) 245–252. https://doi.org/10.1287/trsc.18.3.245.

[34] Xiaofan Wang, Xiang Li, Guanrong Chen, Basic topological properties of networks, in: Introduction to Network Science (in Chinese), Higher Education Press, B

eijing China, 2012: pp. 104–112.

[35] H. Hu, X. Wang, Unified index to quantifying heterogeneity of complex networks., Physica A: Statistical Mechanics and Its Applications 387 (2008) 3769–3780. https://doi.org/10.1016/j.physa.2008.01.113.

[36] G. Yan, N.D. Martinez, Y.-Y. Liu, Degree heterogeneity and stability of ecological networks, J R Soc Interface 14 (2017) 1–10. https://doi.org/10.1098/rsif.2017.0189.

[37] M. Newman, The configuration model, in: Networks, 2nd Edition, Oxford University Press, 2018: pp. 369–443. https://doi.org/10.1093/oso/9780198805090.001.0001.

[38] S.A. Ozbay, M.M. Nguyen, Parameterizing network graph heterogeneity using a modified Weibull distribution, Appl. Network Sci. 8 (2023) 20. https://doi.org/10.1007/s41109-023-00544-9.

## Appendix

**Appendix A**

Solving for $x_C$ in **Eq.** (18):

$$\dot{x}_C = \begin{cases} -x_C \cdot \dfrac{2(k-r)}{k} \cdot cost & , r < k \\ 0 & , r = k \\ (1-x_C)\dfrac{2(r-k)}{k} \cdot cost & , r > k \end{cases}$$

we solve for $x_C$ when $r > k$. Let $\eta = \dfrac{2(r-k)}{k} \cdot cost$ and the initial value of $x_C$ be $x_0$. Thus, $\eta > 0$, $x_0 \in [0,1]$, we obtain the following expression:

$$\begin{aligned} dx_C / dt &= \eta(1-x_C) \\ dx_C / (1-x_C) &= \eta dt \\ \int \frac{1}{1-x_C} dx_C &= \int \eta dt \\ -\ln|1-x_C| &= \eta t + C_1 \\ x_C &= 1 \pm C_2 e^{-\eta t} \end{aligned}$$

Since $x_C \leq 1$, we deduce that $x_C = 1 - C_2 e^{-\eta t}$. Substituting $t = 0$, $x_C = x_0$ into the expression, we arrive at $C_2 = 1 - x_0$. Therefore,

$$x_C = 1-(1-x_0)e^{-\eta t} = 1-(1-x_0)e^{-\frac{2(r-k)\cdot cost}{k}t}$$

## Appendix B

The original **Table 1** is discussed in cases based on the relationship between $r$, $k$, and $\theta$. According to the conditions $1 < r \leq k/\theta$, $k/\theta < r \leq k$, $k < r \leq 2k$, $r > 2k$, Tables S1 to S4 are organized as follows:

**Table S1** When $1 < r \leq k/\theta$, expressions for $\tau_{D \to C}(e_m)$ and $\tau_{C \to D}(e_m)$ under threshold constraints

| | \multicolumn{4}{c}{$\tau_{D \to C}(e_m)$} | | | |
|---|---|---|---|---|
| | $n_p \leq \theta - 1$ | $n_p = \theta$ | $n_p = \theta + 1$ | $n_p > \theta + 1$ |
| $n_q \leq \theta - 1$ | 0 | 0 | 0 | 0 |
| $n_q = \theta$ | 0 | 0 | 0 | 0 |
| $n_q = \theta + 1$ | 0 | 0 | 0 | 0 |
| $n_q > \theta + 1$ | 0 | 0 | 0 | 0 |
| | \multicolumn{4}{c}{$\tau_{C \to D}(e_m)$} | | | |
| | $n_p \leq \theta - 1$ | $n_p = \theta$ | $n_p = \theta + 1$ | $n_p > \theta + 1$ |
| $n_q \leq \theta - 1$ | $\frac{2(k-r)}{k} \cdot cost$ | $\frac{2(k-r)}{k} \cdot cost$ | $\frac{2k - r(1-\theta)}{k} \cdot cost$ | $\frac{2(k-r/2)}{k} \cdot cost$ |
| $n_q = \theta$ | $\frac{2(k-r)}{k} \cdot cost$ | $\frac{2(k-r)}{k} \cdot cost$ | $\frac{2k - r(1-\theta)}{k} \cdot cost$ | $\frac{2(k-r/2)}{k} \cdot cost$ |
| $n_q = \theta + 1$ | $\frac{2k - r(1-\theta)}{k} \cdot cost$ | $\frac{2k - r(1-\theta)}{k} \cdot cost$ | $\frac{2(r\theta + k)}{k} \cdot cost$ | $\frac{2(r\theta/2 + k)}{k} \cdot cost$ |
| $n_q > \theta + 1$ | $\frac{2(k-r/2)}{k} \cdot cost$ | $\frac{2(k-r/2)}{k} \cdot cost$ | $\frac{2(r\theta/2 + k)}{k} \cdot cost$ | $2cost$ |

**Table S2** When $k/\theta < r \leq k$, expressions for $\tau_{D \to C}(e_m)$ and $\tau_{C \to D}(e_m)$ under threshold constraints

| | \multicolumn{4}{c}{$\tau_{D \to C}(e_m)$} | | | |
|---|---|---|---|---|
| | $n_p \leq \theta - 1$ | $n_p = \theta$ | $n_p = \theta + 1$ | $n_p > \theta + 1$ |
| $n_q \leq \theta - 1$ | 0 | 0 | 0 | 0 |
| $n_q = \theta$ | 0 | $\frac{2(r\theta - k)}{k} \cdot cost$ | 0 | 0 |
| $n_q = \theta + 1$ | 0 | 0 | 0 | 0 |
| $n_q > \theta + 1$ | 0 | 0 | 0 | 0 |
| | \multicolumn{4}{c}{$\tau_{C \to D}(e_m)$} | | | |
| | $n_p \leq \theta - 1$ | $n_p = \theta$ | $n_p = \theta + 1$ | $n_p > \theta + 1$ |
| $n_q \leq \theta - 1$ | $\frac{2(k-r)}{k} \cdot cost$ | $\frac{2(k-r)}{k} \cdot cost$ | $\frac{2k - r(1-\theta)}{k} \cdot cost$ | $\frac{2(k-r/2)}{k} \cdot cost$ |
| $n_q = \theta$ | $\frac{2(k-r)}{k} \cdot cost$ | $\frac{2(k-r)}{k} \cdot cost$ | $\frac{2k - r(1-\theta)}{k} \cdot cost$ | $\frac{2(k-r/2)}{k} \cdot cost$ |
| $n_q = \theta + 1$ | $\frac{2k - r(1-\theta)}{k} \cdot cost$ | $\frac{2k - r(1-\theta)}{k} \cdot cost$ | $\frac{2(r\theta + k)}{k} \cdot cost$ | $\frac{2(r\theta/2 + k)}{k} \cdot cost$ |
| $n_q > \theta+1$ | $\frac{2(k-r/2)}{k} \cdot cost$ | $\frac{2(k-r/2)}{k} \cdot cost$ | $\frac{2(r\theta/2+k)}{k} \cdot cost$ | $2cost$ |

**Table S3** When $k < r \leq 2k$, expressions for $\tau_{D \to C}(e_m)$ and $\tau_{C \to D}(e_m)$ under threshold constraints

| $\tau_{D \to C}(e_m)$ | | | | |
|---|---|---|---|---|
| | $n_p \leq \theta-1$ | $n_p = \theta$ | $n_p = \theta+1$ | $n_p > \theta+1$ |
| $n_q \leq \theta-1$ | $\frac{2(r-k)}{k} \cdot cost$ | 0 | 0 | 0 |
| $n_q = \theta$ | 0 | $\frac{2(r\theta-k)}{k} \cdot cost$ | 0 | 0 |
| $n_q = \theta+1$ | 0 | 0 | 0 | 0 |
| $n_q > \theta+1$ | 0 | 0 | 0 | 0 |

| $\tau_{C \to D}(e_m)$ | | | | |
|---|---|---|---|---|
| | $n_p \leq \theta-1$ | $n_p = \theta$ | $n_p = \theta+1$ | $n_p > \theta+1$ |
| $n_q \leq \theta-1$ | 0 | 0 | $\frac{2k-r(1-\theta)}{k} \cdot cost$ | $\frac{2(k-r/2)}{k} \cdot cost$ |
| $n_q = \theta$ | 0 | 0 | $\frac{2k-r(1-\theta)}{k} \cdot cost$ | $\frac{2(k-r/2)}{k} \cdot cost$ |
| $n_q = \theta+1$ | $\frac{2k-r(1-\theta)}{k} \cdot cost$ | $\frac{2k-r(1-\theta)}{k} \cdot cost$ | $\frac{2(r\theta+k)}{k} \cdot cost$ | $\frac{2(r\theta/2+k)}{k} \cdot cost$ |
| $n_q > \theta+1$ | $\frac{2(k-r/2)}{k} \cdot cost$ | $\frac{2(k-r/2)}{k} \cdot cost$ | $\frac{2(r\theta/2+k)}{k} \cdot cost$ | $2cost$ |

**Table S4** When $r > 2k$, expressions for $\tau_{D \to C}(e_m)$ and $\tau_{C \to D}(e_m)$ under threshold constraints

| $\tau_{D \to C}(e_m)$ | | | | |
|---|---|---|---|---|
| | $n_p \leq \theta-1$ | $n_p = \theta$ | $n_p = \theta+1$ | $n_p > \theta+1$ |
| $n_q \leq \theta-1$ | $\frac{2(r-k)}{k} \cdot cost$ | 0 | $\frac{2(r/2-k)}{k} \cdot cost$ | $\frac{2(r/2-k)}{k} \cdot cost$ |
| $n_q = \theta$ | 0 | $\frac{2(r\theta-k)}{k} \cdot cost$ | 0 | 0 |
| $n_q = \theta+1$ | $\frac{2(r/2-k)}{k} \cdot cost$ | 0 | 0 | 0 |
| $n_q > \theta+1$ | $\frac{2(r/2-k)}{k} \cdot cost$ | 0 | 0 | 0 |

| $\tau_{C \to D}(e_m)$ | | | | |
|---|---|---|---|---|
| | $n_p \leq \theta-1$ | $n_p = \theta$ | $n_p = \theta+1$ | $n_p > \theta+1$ |
| $n_q \leq \theta-1$ | 0 | 0 | $\frac{2k-r(1-\theta)}{k} \cdot cost$ | 0 |
| $n_q = \theta$ | 0 | 0 | $\frac{2k-r(1-\theta)}{k} \cdot cost$ | 0 |
| $n_q = \theta+1$ | $\frac{2k-r(1-\theta)}{k} \cdot cost$ | $\frac{2k-r(1-\theta)}{k} \cdot cost$ | $\frac{2(r\theta+k)}{k} \cdot cost$ | $\frac{2(r\theta/2+k)}{k} \cdot cost$ |
| $n_q > \theta+1$ | 0 | 0 | $\frac{2(r\theta/2+k)}{k} \cdot cost$ | $2cost$ |

**Appendix C**

**Fig.S1** presents the proportion of cooperators (blue solid line, $y_1$), the group average payoff (red solid line, $y_2$), and the number of nodes exceeding the threshold constraint (green solid line, $y_3$) as functions of the initial number of cooperators ($M \cdot x_0$) for five types of homogeneous network evolution states at different values of $r$. As $M \cdot x_0$ changes, we observe that stable high cooperation proportions can be maintained. Specifically, in subplots (**k**) to (**m**) and (**p**) to (**r**), stable cooperation (with $y_1 \approx 0.48$) is observed for *Network* 1 ~ 3, where $r = k_{max} + 0.1$ and $r = 2k_{min} - 0.1$. For other ranges of $r$ values and network types, the proportion of cooperators either decreases or enters the "excessive cooperation" state ($y_3 > 0$). Notably, when network heterogeneity increases (*Network* 4 ~ 5), the condition $2k_{min} \leq k_{max}$, $r > k_{max}$ leads to the occurrence of $y_3 > 0$ as the initial cooperation rate increases. Furthermore, the larger the value of $r$, the smaller the $M \cdot x_0$ value at which this occurs. When $r \leq 2k_{min}$, the cooperation rate $y_1$ shows a decline and is noticeably lower than the ideal cooperation structure. These results indicate that networks with high heterogeneity are difficult to evolve into the optimal cooperation structure by merely adjusting a single $r$.

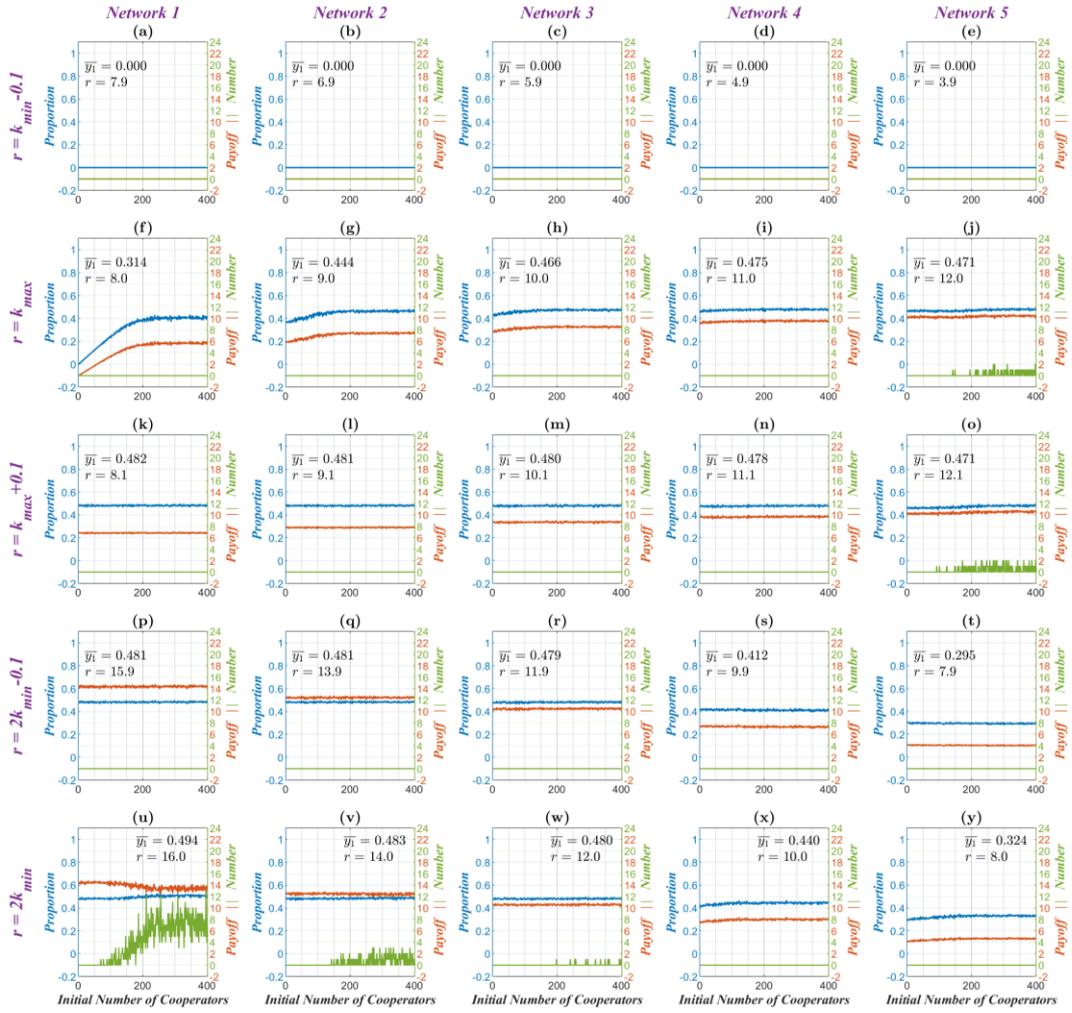

**Fig.S1** Under threshold constraints, the evolution of the cooperator proportion (blue line), average group payoff (orange line), and the number of nodes exceeding the threshold (green line) for five homogeneous networks under five different $r$-values, as a function of the initial cooperator quantity (**a ~ y**).

**Fig.S2** compares the evolutionary equilibrium of cooperation in networks with 8 different node and edge configurations, derived from the Weibull probability density distribution, under initial cooperation proportions of 0 and 1. The figure shows the changes in the cooperation proportion (blue line), the average group payoff (orange line), and the number of exceeding threshold constraint nodes (green line) as the value of $r$ changes. *Network* I~VIII correspond to increasing degrees of heterogeneity in network degree. Panels (**a ~ d**) and (**i ~ l**) correspond to the case where the initial cooperation proportion is 0, while panels (**e ~ h**) and (**m ~ p**) correspond to the case where the initial cooperation proportion is 1. The results for *Network* I~III are similar to those in **Fig. 14** of the original text and **Fig.S1**. When the degree heterogeneity of the network increases to σ ≥ 0.37 (**IV~VIII**), regardless of the initial cooperation proportion being 0 or 1, it becomes impossible to find a fixed *r* value that satisfies the "moderate cooperation" condition. This further confirms that networks with high heterogeneity cannot achieve an optimal cooperative structure through evolution under a fixed *r* value.

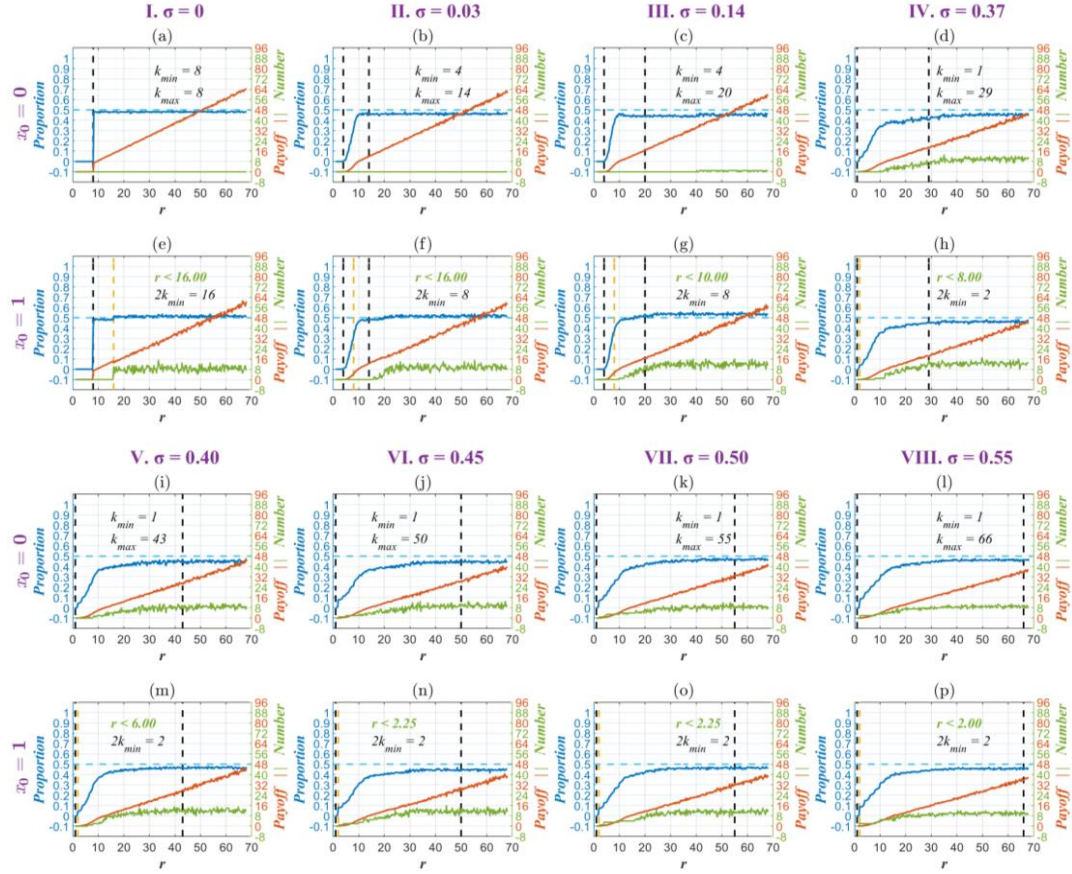

**Fig.S2** Comparison of the cooperator proportion (blue line), average group payoff (orange line), and the number of nodes exceeding the threshold (green line) for networks with the same number of nodes and edges, based on the Weibull probability density distribution. The results are shown as a function of the *r*-value. *Network*s I~VIII correspond to increasing degree heterogeneity. In panels (**a ~ d**) and (**i ~ l**), the initial cooperator proportion is 0, and in panels (**e ~ h**) and (**m ~ p**), the initial cooperator proportion is 1. The black dashed lines represent reference lines for $r = k_{min}$ and $r = k_{max}$, while the yellow dashed line represents the reference line for $r = 2k_{min}$, and the light blue dashed line represents the reference line for $y_1 = \theta/k$.

**Fig.S3** compares the proportion of cooperators (panels **a, d**), average group payoff (panels **b, e**), and the number of nodes exceeding the threshold constraints (panels **c, f**) at evolutionary equilibrium for eight networks with the same number of nodes and edges, based on the Weibull probability density distribution, after 100 replicate experiments with different *n*-fold values (0.9, 1, 1.1, 1.9, 2, and 2.1) for initial cooperator proportions of 0 and 1. The findings are as follows:

(1) When $x_0 = 0$: For *n-fold* values of 0.9 and 1, $y_1 = 0$, $y_2 = 0$. For other *n-fold* values, the results for $y_1$ are similar across the same network, and although the average payoff $y_2$ varies in magnitude, the trend remains the same. In all cases, $y_3 = 0$. This indicates that when $x_0 = 0$, *n-fold* > 1 can achieve "moderate cooperation", and as network heterogeneity increases, the cooperator proportion of the ESS shows a declining trend.

(2) When $x_0 = 1$: For *n-fold* values of 0.9, $y_1 = 0$, $y_2 = 0$, and $y_3 = 0$. For *n-fold* = 1, differences emerge. $y_1$ and $y_2$ are no longer zero and start to show positive values, though they are smaller than those for other *n-fold* values, and $y_3$ remains zero. For *n-fold* = 1.1 and 1.9, the $y_1$ curves almost overlap, and both $y_1$ and $y_2$ curves are similar to those seen when $x_0 = 0$, with $y_3=0$, indicating "moderate cooperation". For *n-fold* values of 2 and 2.1, $y_1$ increases dramatically, and the $y_2$ curve is notably lower than for *n-fold* values of 1.1 and 1.9. As network heterogeneity increases, the difference grows larger, and $y_3$ is no longer zero. This suggests that when $x_0 = 1$, "moderate cooperation" can be achieved only when *n-fold* is between 1 and 2. For *n-fold* ≥ 2, the system enters the "excessive cooperation" mode.

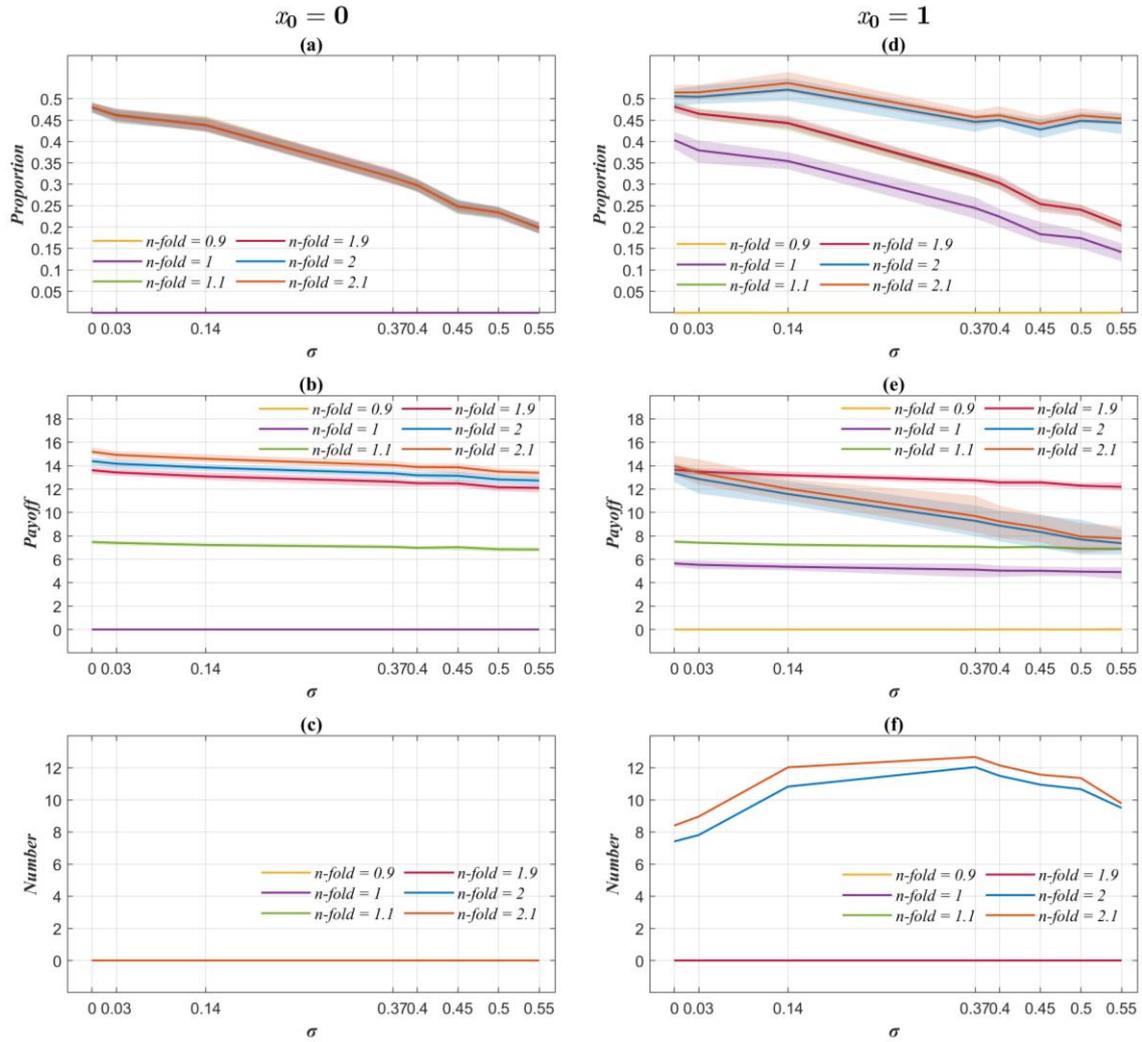

**Fig.S3** Under threshold constraints, the cooperator proportion (**a, d**), average group payoff (**b, e**), and the number of nodes exceeding the threshold (**c, f**) for heterogeneous networks with the same number of nodes and edges, based on the Weibull probability density distribution, are shown as a function of the heterogeneity parameter $\sigma$, for six different *n-fold* values.